\documentclass[useAMS,usenatbib,usegraphicx]{mn2e}
\usepackage{mathptmx}
\usepackage{aas_macros}
\usepackage{amsmath}
\usepackage{amssymb}
\usepackage{mathtools}
\usepackage{graphicx}
\newcommand{\beq}{\begin{equation}}
\newcommand{\eeq}{\end{equation}}

\usepackage{txfonts}

\title[Co-orbiting satellite planes]{Co-orbiting satellite galaxy structures are still in conflict with the distribution of primordial dwarf galaxies}
\author[Pawlowski et al.]{Marcel S. Pawlowski$^{1}$\thanks{E-mail:
marcel.pawlowski@case.edu}, 
Benoit Famaey$^{2}$, 
Helmut Jerjen$^{3}$,
David Merritt$^{4}$,
Pavel Kroupa$^{5}$,
\newauthor
J\"org Dabringhausen$^{6}$,
Fabian L\"ughausen$^{5}$,
Duncan A. Forbes$^{7}$,
Gerhard Hensler$^{8}$,
\newauthor
Fran\c{c}ois Hammer$^{9}$,
Mathieu Puech$^{9}$,
Sylvain Fouquet$^{9}$,
Hector Flores$^{9}$,
Yanbin Yang$^{9}$\\
$^{1}$Department of Astronomy, Case Western Reserve University, 10900 Euclid Avenue, Cleveland, OH 44106, USA\\
$^{2}$Observatoire astronomique de Strasbourg, Universit\'e de Strasbourg, CNRS, UMR 7550, 11 rue de l'Universit\'e, F-67000 Strasbourg, France\\
$^{3}$Research School of Astronomy and Astrophysics, Australian National University, Mt Stromlo Observatory, Cotter Rd., Weston ACT 2611, Australia\\
$^{4}$School of Physics and Astronomy and Center for Computational Relativity and Gravitation, Rochester Institute of Technology, 84 Lomb Memorial Drive,\\ Rochester, NY 14623, USA\\
$^{5}$Helmholtz-Institut f\"ur Strahlen- und Kernphysik, Rheinische Friedrich-Wilhelms-Universit\"at Bonn, Nussallee 14-16, D-53115 Bonn, Germany\\
$^{6}$Departamento de Astronom\'ia, Universidad de Concepci\'on, Casilla 160-C, Concepci\'on, Chile\\
$^{7}$Centre for Astrophysics \& Supercomputing, Swinburne University, Hawthorn VIC 3122, Australia\\
$^{8}$Department of Astrophysics, University of Vienna, T\"urkenschanzstr. 17, 1180 Vienna, Austria\\
$^{9}$Laboratoire GEPI, Observatoire de Paris, CNRS-UMR8111, Univ. Paris Diderot, 5 place Jules Janssen, F-92195 Meudon, France
}
\begin{document}
\date{Accepted 2014 May 18.  Received 2014 April 11; in original form 2014 February 3}
\pagerange{\pageref{firstpage}--\pageref{lastpage}} \pubyear{2014}
\maketitle
\label{firstpage}
\begin{abstract}
Both major galaxies in the Local Group host planar distributions of co-orbiting satellite galaxies, the Vast Polar Structure (VPOS) of the Milky Way and the Great Plane of Andromeda (GPoA). The $\Lambda$CDM cosmological model did not predict these features. However, according to three recent studies the properties of the GPoA and the flattening of the VPOS are common features among sub-halo based $\Lambda$CDM satellite systems, and the GPoA can be naturally explained by satellites being acquired along cold gas streams. We point out some methodological issues in these studies: either the selection of model satellites is different from that of the observed ones, or an incomplete set of observational constraints has been considered, or the observed satellite distribution is inconsistent with basic assumptions. Once these issues have been addressed, the conclusions are different: features like the VPOS and GPoA are very rare (each with probability $\lesssim 10^{-3}$, and combined probability $< 10^{-5}$) if satellites are selected from a $\Lambda$CDM simulation combined with semi-analytic modelling, and accretion along cold streams is no natural explanation of the GPoA. 
The origin of planar dwarf galaxy structures remains unexplained in the standard paradigm of galaxy formation.
\end{abstract}

\begin{keywords}
Galaxies: dwarf -- Galaxies: kinematics and dynamics -- Local Group -- dark matter -- methods: data analysis
\end{keywords}

%________________________________________________________________

\section{Introduction}

In recent years the notion that satellite objects of the Milky Way (MW) are preferentially distributed and orbit within a common plane \citep{Kunkel1976,LyndenBell1976}, a vast polar structure (VPOS)\footnote{The feature which we call the VPOS is called the 'Magellanic plane' by \citet{Kunkel1976} and \citet{LyndenBell1976}; the 'great pancake' by \citet{Libeskind2005}, \citet{Li2008} and \citet{Wang2013}; and the disc of satellites by \citet{Kroupa2005} and \citet{Metz2008}.}, has been substantiated by additional and improved observational information. The distribution of the MW satellite galaxies can be described by a relatively thin (about $20 \,$kpc root-mean-square height) plane \citep{Kroupa2005,Metz2007,Pawlowski2013a}, whilst their proper motions indicate that up to 9 out of the 11 brightest (termed the ``classical'' ones by \citealt{Metz2008}) MW satellites orbit in the same sense (``co-orbit'') within the structure \citep{Metz2008,Pawlowski2013b}. Globular clusters and stellar and gaseous streams appear to preferentially align with the VPOS, too \citep{Forbes2009,Pawlowski2012a,Keller2012}.

A similar planar distribution of satellites has been sought around M31 \citep{McConnachie2006,Koch2006,Metz2007}, but it required the homogeneous dataset of M31 satellite distances provided by \citet{Conn2012} to allow \citet{Ibata2013} and \citet{Conn2013} to discover such a thin (of about $10\,$kpc root-mean-square height) planar satellite structure with high statistical significance. Like the VPOS, this Great Plane of Andromeda (GPoA)\footnote{The same structure has been called the vast thin disk of satellites (VTDS) by \citet{Hammer2013} and the vast thin plane of dwarf galaxies (VTPD) by \citet{Bahl2014}.} appears to be co-orbiting, but it consists of only about half the currently known M31 satellites. Similar alignments of dwarf galaxies have since been identified among the secluded dwarf galaxies in the Local Group (LG) \citep{Pawlowski2013a,Bellazzini2013,Pawlowski2014} and aligned systems of satellites and stellar streams are also being discovered around more distant galaxies \citep{Galianni2010,Duc2011,Paudel2013,Karachentsev2014}.

The existence of the thin, planar VPOS is a challenge for the standard $\Lambda$-cold dark matter cosmological model (the $\Lambda$CDM model). \citet{Kroupa2005} noted for the first time that the distribution of dark matter sub-haloes in cosmological simulations is isotropic to first order, thus being incompatible with the planar VPOS. This realization found further support in the fact that most of the luminous MW satellites orbit in the same direction within the VPOS \citep{Metz2008,Pawlowski2013b}.

The phase-space correlation of both satellite galaxy structures (VPOS and GPoA) has led to the suggestion \citep{Metz2007b} that a large fraction of the observed satellites of the MW and of M31 are in fact tidal dwarf galaxies \citep[TDGs, e.g.][]{Barnes1992,Bournaud2008} formed in the debris shed by a past major galaxy encounter rather than being dark-matter dominated primordial dwarf galaxies. The formation of TDGs out of the debris of galaxy collisions has been observed to take place even in the current Universe and it can be considered observationally and numerically confirmed that this process results in phase-space correlated satellite systems \citep[e.g.][]{Wetzstein2007,Bournaud2007,Bournaud2008,Duc2011,Duc2014}. TDGs have been found to be indistinguishable from dwarf elliptical galaxies \citep{Dabringhausen2013}, are able to survive their initial burst of star formation \citep{Recchi2007,Ploeckinger2014}, and might also be responsible for the larger-scale dwarf galaxy planes in the LG \citep{Pawlowski2013a,Bellazzini2013,Pawlowski2014}.

A number of possible galaxy encounter scenarios in the LG have been suggested to explain the formation of the observed dwarf galaxy structures, such as the tidal disruption of a LMC-progenitor \citep{Kunkel1976,LyndenBell1976,Pawlowski2011} and a past encounter between the MW and M31 \citep{Sawa2005,Pawlowski2012a,Zhao2013}. The currently most advanced scenario of such an event is that of a major merger in M31 by \citet{Hammer2010}. Their numerical models were initially tailored to reproduce features observed in M31, but also resulted in the formation of TDGs from the tidal debris. Some of these turned out to reproduce the later observed GPoA \citep{Hammer2013}, while others are flung out towards the MW where they form a VPOS-like structure \citep{Yang2010,Fouquet2012}, which is consistent with the alignment of the GPoA plane towards the MW.

The virtue of the suggested TDG-origin is that, for a given scenario, it is possible to make detailed predictions of the orbits of TDGs formed in such an encounter. This makes the different scenarios of TDG formation testable, in contrast to the scenario in which the satellites have been accreted as primordial dwarf galaxies in an essentially stochastic manner, which results in an inherent inability to predict their exact orbits.
However, since being identified as a potential problem for the $\Lambda$CDM cosmological model, numerous attempts have been made to reconcile the observed satellite structures with a cosmological origin.

A proposed scenario is the accretion of groups of satellite galaxies, which would naturally result in them having similar orbital parameters \citep{Donghia2008,Li2008}. However, \citet{Metz2009} have shown that this explanation is inconsistent with observed properties of dwarf galaxy groups. They are too extended to form narrow structures such as the VPOS or the GPoA, both of which have a thickness of $\lesssim 50$\,kpc.

\citet{Lovell2011} claimed that quasi-planar distributions of coherently rotating satellites similar to the VPOS arise naturally in simulations of a $\Lambda$CDM universe. However, using their data \citet{Pawlowski2012b} tested how often the orbital poles (directions of angular momenta) of sub-haloes are as concentrated and as perpendicular to the Galactic disc normal as those of the observed MW satellites. The authors were unable to find support for the conclusion by \citet{Lovell2011}. Instead, an orbital pole distribution similar to the observed one is extremely unlikely if the satellite orbital poles are drawn from those of cosmological sub-haloes, but very likely if the orbital poles are drawn from those of tidal debris, as would be the case for TDG-satellites.

\citet{Libeskind2009} and \citet{Deason2011} are two of the few studies that in addition to using the spatial information of dark matter sub-haloes from cosmological simulations also compare the orbital poles of sub-haloes with those of the observed satellite galaxies. \citet{Libeskind2009} find that in approximately 35 per cent of the simulated satellite galaxy systems they consider, the orbital poles of at least 3 out of 11 satellite galaxies point along the short axis of the galaxy distribution. \citet{Deason2011}, searching for alignments of 3 out of only 10 satellite galaxies, find this to be the case for approximately 20 per cent of their simulated systems.
However, as we have shown in \citet{Pawlowski2013b}, additional and improved proper motion data for the 11 brightest MW satellites results in a much more concentrated distribution of orbital poles for the co-orbiting satellites, such that now at least six of the orbital poles are well aligned. This is an extremely rare finding for $\Lambda$CDM-based satellites according to fig. 9 of \citet{Libeskind2009} and fig. 8 of \citet{Deason2011}.

Three new attempts at reconciling the observed GPoA and VPOS with a purely cosmological origin have been made in the past two years. These studies claim that: 
(1) structures such as the GPoA in M31 are not rare among satellite systems in the Millennium II simulation and thus the GPoA is consistent with $\Lambda$CDM \citep[][BB14]{Bahl2014}, 
(2) structures such as the VPOS in the MW are in agreement with a $\Lambda$CDM universe because flattened satellite distributions as the one around the MW are frequent in the Millennium II simulation \citep[][W13]{Wang2013},
(3) the GPoA can be naturally explained if host galaxies acquire not only their gas but also their satellites along cold streams and that the GPoA can therefore be interpreted as indirect observational evidence of the paradigm of cold mode accretion streams \citep[][GB13]{Goerdt2013}.

When searching for GPoA or VPOS analogues in cosmological simulations, it is important to decide which are their fundamental parameters. The most obvious ones describe the overall spatial distribution. For example the root-mean-square (rms) height of the satellites around their best-fitting plane ($r_{\mathrm{per}}$) and their radial distribution in that plane measured by the rms of their distances projected into the plane ($r_{\mathrm{par}}$). An alternative is using the relative shape measured by the rms axis ratios ($c/a$\ and $b/a$). 
But the satellite planes are structures in six-dimensional phase space. The VPOS is rotationally stabilized \citep{Metz2007,Fouquet2012,Pawlowski2013b} and the coherent velocity directions of the satellites belonging to the GPoA were, in fact, the most important part of the discovery of the structure \citep{Ibata2013}. When comparing simulations to observations, an analysis would therefore be incomplete if it would focus on only part of the observable dimensions such as part of the position space by only considering $r_{\mathrm{per}}$, in particular when velocities have been used observationally to detect the structures \citep[e.g.][]{Metz2008,Libeskind2009,Deason2011,Pawlowski2012b,Pawlowski2013b,Ibata2013} and are available from the simulation.

The minimum required parameters can thus be summarized by trying to give the shortest possible characterisation of our current knowledge of the two satellite structures in the LG:
(1) The GPoA is thin ($r_{\mathrm{per}}$) but extended ($r_{\mathrm{par}}$) and has coherent velocity directions ($n_{\mathrm{coorb}}$, number of satellites showing the same velocity trend).
(2) The VPOS is thin ($r_{\mathrm{per}}$) but extended ($r_{\mathrm{par}}$) and the satellite orbital poles are strongly concentrated ($\Delta_{\mathrm{std}}$, spherical standard distance of the concentrated orbital poles) in the direction of the plane normal ($\theta_{\mathrm{VPOS}}$, angle between plane normal direction and average direction of concentrated orbital poles).
In order to be as conservative as possible these parameters will only be considered as limits in the following. For example, the rms height of a simulated satellite plane is not required to match the observed values closely, but only to be smaller or equal to the observed value.

This paper is structured as follows. In Sect. \ref{sect:bahl} we search for GPoA-like planes in the same simulation dataset used by BB14. We use the same simulation dataset in Sect. \ref{sect:wang} to search for VPOS-like satellite planes, as in the study by W13. In Sect. \ref{sect:goerdt} we test whether cold accretion streams can naturally explain the GPoA, as claimed by GB13. We close the paper with concluding remarks in Sect. \ref{sect:conclusion}.

Sections \ref{sect:bahl} to \ref{sect:goerdt} are all similarly structured. They start with a discussion of the previously applied method and a comparison with the available observational data, followed by the description of the method applied by us. The results of this new method are presented, ending in a summarizing discussion.
For observations, if not mentioned otherwise, in the following we will use the positions of M31 and its satellite galaxies based on \citet{Conn2012}, while all remaining positions and the line-of-sight velocities are based on the catalogue of LG galaxies by \citet{McConnachie2012}. All properties of the simulated satellite systems correspond to the situation at a redshift of $z = 0$.

%__________________________________________________________________

\section{Searching for a M31 GPoA-like structure in $\Lambda$CDM simulations}
\label{sect:bahl}

\subsection{Previous method and its limitations}

As mentioned in \citet{Conn2013}, a satellite galaxy plane such as the GPoA is not expected in a $\Lambda$CDM cosmology. The discussion in \citet{Ibata2013} also illustrates that there is no obvious explanation for this structure. However, both studies have focussed on the observed satellite distribution only. How likely a structure such as the GPoA is in the standard galaxy formation model based on the $\Lambda$CDM cosmology can be quantified by comparing the observed satellite system with simulations of M31-like galaxies. Herein lies the importance of the approach taken by BB14. They have searched for GPoA-like structures among simulated satellite galaxy systems based on the Millennium II simulation \citep{BoylanKolchin2009}. This dark-matter-only simulation has been scaled to match the cosmological parameters of WMAP7 \citep{Jarosik2011} and has then been populated with galaxies using the semi-analytic galaxy formation models by \citet{Guo2013}. For details we refer the reader to these papers.

BB14 conclude that the GPoA is not in conflict with the standard cosmological framework. According to their analysis, satellite configurations as thin as the observed GPoA are quite common around galaxies with a mass comparable to M31 and a significant fraction of the galaxies in the simulated satellite systems are co-orbiting.

A statistical test has to compare consistent samples. This means that the selection applied to the observed satellite system also has to be applied to the simulated one. This might only be possible approximately, but the selection should be made as similar as possible. In addition, it has to be made sure that the selection applied to the simulated satellite system is also applied to the observed satellites before the plane parameters are determined for both, preferentially by using the same method. Only then can a comparison of the plane parameters yield meaningful results. The analysis of BB14 is, in this respect, limited. We have identified the following problems:
\begin{enumerate}
  \item Three criteria were introduced which must be met by a satellite system to be similar to the observed GPoA, based on the properties of the best-fitting (GPoA-like) plane consisting of a sub-set of the satellite galaxies (15 out of 27 satellites in BB14). These properties are (1) the root-mean-square (rms) height $r_{\mathrm{per}}$\ of the 15 closest satellites from the best-fitting plane, (2) the rms distance from the centre of the host galaxy, $r_{\mathrm{par}}$, of these 15 satellite galaxies projected into the best-fitting plane and (3) the number of co-orbiting satellites ($n_{\mathrm{coorb}}$). A plane in the simulated satellite systems has to be at least as narrow as the observed one. The plane also has to be at least as radially extended as observed ($r_{\mathrm{par}}^{\mathrm{sim}} \geq r_{\mathrm{par}}^{\mathrm{obs}}$)
, otherwise $r_{\mathrm{per}}$\ might be small because the whole satellite system is very radially concentrated but not necessarily flat. Finally there have to be at least as many co-orbiting satellite galaxies as observed in the GPoA. Despite introducing these three independent criteria, BB14 discuss combinations of only two of them (either $r_{\mathrm{per}}$\ and $n_{\mathrm{coorb}}$\ or $r_{\mathrm{per}}$ and $r_{\mathrm{par}}$). As only a small fraction (2 per cent) of the satellite systems fulfils the two-criteria combinations\footnote{Note that the notion of ``rare" is already here slightly subjective, as a 2\% occurrence could certainly already be labelled as ``not very common".} in the first place, it is important to determine how much lower this fraction drops for systems reproducing all three properties of the observed GPoA.
  \item The PAndAS area was approximated as a cylinder with a radius of 128\,kpc (Bahl, private communication).  Drawing the simulated satellites from a cylindrical volume assumes an infinite distance to the MW, in contrast to the real PAndAS survey which is better approximated by a conical volume with the MW in the apex.
   \item The used method does not guarantee that they only identify (oblate) planes, because narrow prolate distributions of satellites can have small $r_\mathrm{per}$\ too. In addition, they apparently consider a too-small PAndAS footprint of 128\,kpc radius at the distance of M31, while \citet{Ibata2013} and \citet{Conn2013} clearly state that the radius is approximately 150\,kpc. This results in a 27 per cent smaller area from which simulated satellites are selected, which makes the overall satellite distribution more narrow. As prolate distributions are also accepted as planes, this method is biased towards finding narrower structures (having smaller $r_{\mathrm{per}}$) which artificially increases the agreement with the observed situation.
  \item 27 satellites are drawn from an area around M31 that is supposed to model the PAndAs survey footprint. However, while the major part of the PAndAS survey indeed covers an approximately circular area of 150\,kpc radius at the distance of M31, the survey also covers an area of approximately 50\,kpc radius around M33. This is important as two of the 27 M31 satellites covered by PAndAs are outside the 150\,kpc-zone: M33 and Andromeda XXII. Both objects are not associated with the GPoA. An alternative to using a circular footprint would have been to use the exact PAndAs footprint from different random viewing angles \citep{Ibata2014}. However, if one chooses to stay as close as possible to the original method of BB14, i.e. simulated satellites are only drawn from within the 150\,kpc-zone, then only 25 satellites must be drawn from the simulation. Therefore, instead of searching for planar alignments of 15 out of 27 satellites, BB14 would have had to search for alignments of 15 out of 25. Using a too-large number of simulated satellites increases the probability to find narrow alignments of satellites because more objects are available to select from in the same volume. 
  As an example, the number of possible combinations of 15 satellites out of 27 is $\frac{27!}{15! 12!} \approx 17.4 \times 10^{6}$, but only $\frac{25!}{15! 10!} \approx 3.3 \times 10^{6}$\ for 15 out of 25 satellites. Additional satellites therefore substantially increase the chance of finding a more narrow plane if the number of satellites assigned to the plane is kept fixed.
  The choice of the satellite number by BB14 therefore biases the results towards finding a better agreement with the observed situation.
  \item Simulated satellite galaxies were selected from cubical volumes extracted from the Millennium II simulation data which are centred on the host galaxies (Bahl, private communication). These cubes have an edge-length of $600/h$\,kpc, where $h = 0.704$. 
When searching for anisotropies in the simulated satellite systems, it is crucial to draw them from a volume that is non-directional, i.e. a sphere. Otherwise, the 3D geometry of the volume from which satellites are drawn will introduce anisotropies in the data.
  The maximum spherical volume fitting into these cubes has a radius of $r_{\mathrm{sphere}} = 426$\,kpc.
   However, according to figures 3 and 7 in BB14, some of their simulated satellites have distances of 700\,kpc or more from their hosts. This indicates that the satellites have not been drawn from a spherical volume, but from the cubes directly, because in that case the random viewing direction will in some cases be along the corners of the cubes, resulting in a maximum distance from the host galaxy of $d_{\mathrm{max}} = \sqrt{3 r_{\mathrm{sphere}}^2} = 738$\,kpc. As the more distant galaxies are only present in the directions of the eight corners of the considered simulation cube, the inclusion of these galaxies artificially introduces a preferred directional alignment to the satellite galaxy distribution. It also results in different radial distributions, because the maximum distance of a satellite from the host depends on the orientation in which the cube is seen. This has to be avoided when trying to determine the chance of finding satellite alignments in a meaningful way.
   \item As pointed out before, for the simulation data used by BB14 only satellite galaxies within a maximum distance of $d_{\mathrm{max}} = 426$\,kpc can be considered in the analysis, because no spatially unbiased data is present for the more distant simulated galaxies. However, Andromeda XVIII is at a distance of $\approx 460$\,kpc from M31 and therefore outside of the volume considered around each of the simulated host galaxies. Consequently, this galaxy must be excluded from the sample of M31 galaxies and the number of satellites drawn from the simulations also has to be reduced by one. As Andromeda XVIII  is not part of the GPoA, the analysis still has to look for planes made up of 15 satellites. However, the M31 satellite galaxy Andromeda XXVII introduces a further complication here. As seen from the MW, it is superimposed on to a stream of stars. This makes a distance determination to the object difficult, such that the discovery paper only reports a lower limit of its distance, placing it at $\approx 760 $\,kpc from the MW or $\approx 90$\,kpc from M31 \citep{Richardson2011}. In contrast to that, the analysis by \citet{Conn2012} places its most-likely heliocentric distance at an upper limit of $\approx 1250$\,kpc or $\approx 480$\,kpc from M31, but its distance posteriori distribution (see their fig. 12) is strongly bimodal, indicating that the object is either at $\approx 1250$\,kpc heliocentric distance or at $\approx 800$\,kpc. If Andromeda XXVII is near, it falls within the volume extracted from the simulation, if it is far it also has to be excluded from the sample. In summary, to be compatible to the observed distribution of M31 satellites as used in the analysis of \citet{Ibata2013}, one is only allowed to draw 23 satellite galaxies from the simulation data used by BB14 and to search for planar alignments of 14 of these satellites. Assuming a closer distance of Andromeda XXVII increases the numbers by one: 15 satellites out of 24. In both cases the properties of the GPoA plane-fit need to be re-determined as either the satellite sample or the distance position of one of the satellites is different from that used in the analysis of \citet{Ibata2013}.
\end{enumerate}

In summary, the two major issues are that BB14 did not demand all three of the essential GPoA criteria as they were defined in their analysis to be fulfilled simultaneously, and that their selection of model satellites differs from the selection of observed satellites in number, selection area and distance from M31. The importance of the second point can be illustrated as follows: Even if there would be a simulated satellite system in which the 27 satellites with the largest baryonic mass have the exact same positions and velocities as the satellites observed around M31, the analysis of BB14 would not have been able to select all these satellites. The analysis would thus have determined different parameters from a different satellite sample than done observationally by \citet{Ibata2013}, even though the distribution of the top 27 satellites would have been identical.

Finally, we remind the reader of the following other, possibly important aspects which have not been mentioned in
BB14, and which we do not consider either in our new method below:
\begin{enumerate}
\item[(vii)] Their host galaxy sample includes galaxies in high density environments like galaxy clusters, very different environments than where M31 is found. In the future, it might therefore be worthwhile to investigate whether the environment has any influence on the occurrence of co-orbiting satellite galaxy planes.
\item[(viii)] 
The M31 satellite distribution is lopsided, most of the satellite in the GPoA are on the MW-side of M31
\citep{Conn2013}.
\item[(ix)]
The GPoA is orientated almost edge-on towards the MW, being inclined by $\leq 5^{\circ}$\ towards our line-of-sight
\citep{Ibata2013,Conn2013,Pawlowski2013a}.
\end{enumerate}

\subsection{New method}
\label{sect:bahl:method}

\begin{table}
  \caption{Comparison of assumptions.}
  \label{tab:GPoAcomparison}
  \begin{center}  
  \begin{tabular}{lcc}
  \hline
  Parameter & BB14 & this work \\
 \hline
 PAndAS shape & cylinder & cone  \\
 PAndAS radius at host & 128\,kpc & 150\,kpc \\
 Simulation volume shape & cube & sphere \\ 
 Simulation volume size & 852\,kpc edge length & 852\,kpc diameter \\ 
 Andromeda XXVII & far & near or far \\
 $N_{\mathrm{sat}}$ & 27 & 24 or 23 \\
 $N_{\mathrm{GPoA}}$ & 15 & 15 or 14 \\
 \hline
\end{tabular}
 \end{center}
 \small \medskip
Assumptions of the analysis presented by \citet{Bahl2014} (second column) and the ones in this work (third column).
\end{table}

\begin{table*}
 \begin{minipage}{170mm}
  \caption{Parameters of the observed GPoA}
  \label{tab:GPoAparams}
  \begin{center}
  \begin{tabular}{llrr}
  \hline
  Symbol & Description & ``near'' Sample & ``far'' Sample \\
 \hline
$r_\mathrm{M31}^\mathrm{max}$ & Maximum allowed satellite radius from M31 & $300/h$\,kpc & $300/h$\,kpc \\
$d_\mathrm{AndXXVII}$ & Assumed heliocentric distance to Andromeda XXVII & $\approx 800$\,kpc & $\approx 1250$\,kpc \\
$N_{\mathrm{sat}}$ & Number of satellites in sample & 24 & 23 \\
$N_{\mathrm{GPoA}}$ & Number of satellites making up GPoA & 15 & 14 \\
$r_{\mathrm{per}}$ & rms height of GPoA satellites from best-fitting plane & 10.9\,kpc & 10.9\,kpc \\
$r_{\mathrm{par}}$ & rms distance of GPoA satellites projected into best-fitting plane & 147.2\,kpc & 151.0\,kpc \\
$r_{\mathrm{par}}^{\mathrm{all}}$ & rms distance of all $N_{\mathrm{sat}}$ satellites projected into best-fitting plane & 136.8\,kpc & 138.8\,kpc \\
$N_{\mathrm{coorb}}$ & Number of satellites in GPoA which co-orbit & 13 & 13 \\
$c/a$ & Minor to major axis rms height ratio & 0.105 & 0.102 \\
$b/a$ & Intermediate to major axis rms height ratio & 0.788 & 0.740 \\
\hline
$N_{\mathrm{lopside}}$ & Number of GPoA satellites on MW side of M31 & 13 & 13 \\
$\alpha_{\mathrm{MW}}$ & Angle between GPoA and line-of-sight & $4.3^{\circ}$ & $4.9^{\circ}$ \\
\hline
\end{tabular}
 \end{center}
 \small \medskip
The GPoA plane parameters using the observed positions and line-of-sight velocities of the M31 satellite galaxies \citep{Conn2012,McConnachie2012}. They have been determined using the same method applied to the simulated satellite systems. The first column gives the symbol of the quantity described in the second column. The third column assumes that Andromeda XXVII is close to M31, such that it is within the volume covered by the simulated satellite systems. The fourth column assumes that it is outside and should therefore not be considered in the determination of the GPoA parameters that will be compared to the simulated systems.
\end{minipage}
\end{table*}

For best comparability, we use the same data extracted from the Millennium II database \citep{Lemson2006} as was used by BB14. It consists of all galaxies within cubes of edge-length $d_{\mathrm{edge}} = 600/h\,\mathrm{kpc} = 852$\,kpc centered on the main halo in the virial mass range of 1.1 to $1.7 \times 10^{12}\,\mathrm{M}_{\sun}$. There are 1825 such main haloes. Following the recipe in BB14 we remove all main haloes that contain a galaxy with a baryonic mass larger than $7 \times 10^{10} \mathrm{M}_{\sun}$\ within 500\,kpc. Note that the simulation boxes do not reach out to 500\,kpc, such that we can only be sure to remove galaxies within a sphere of radius $d_{\mathrm{edge}}/2$. We then try to reproduce their results, whilst also addressing the caveats mentioned above \footnote{However, as we stick with the exact same simulation dataset, we do not address the potential effect of selecting galaxies residing in clusters rather than completely isolated groups.}. Table \ref{tab:GPoAcomparison} compares our assumptions with those of BB14.

For each host galaxy we construct a sample of satellite galaxies. We only consider satellites with masses above $2.8 \times 10^4\,\mathrm{M}_{\sun}$, again following the cuts by BB14.  
In the semi-analytic galaxy formation model which was applied to the Millennium II simulation, galaxies within sub-haloes that drop below the resolution limit are allowed to survive. The position of these ``orphan'' galaxies is determined by following that particle from the original sub-halo which was most-bound at the last time the sub-halo was resolved. The positions and velocities of these objects are therefore less reliable than those of galaxies embedded in resolved sub-haloes (W13,\citealt{Pawlowski2013a}), but for simplicity and comparability to BB14, as well as W13 (see next Section) we do not discriminate between orphan and non-orphan galaxies.

For each halo a random viewing direction is determined. Then the satellite galaxies with the largest baryonic masses (i.e. the brightest ones) are chosen following four different sample prescriptions:
\begin{enumerate}
   \item ``BB14-like'' sample: $N_{\mathrm{sat}} = 27$\ satellites within a cylinder of 128\,kpc radius and outside of a cylinder of 32\,kpc radius centred on the host galaxy are drawn from the whole simulation cube. There is no radial cut applied and thus satellites can have radii larger than $d_{\mathrm{edge}}/2$\ if the viewing direction is along the corners of the cube. 
   This sample reproduces the sample by BB14, and is included for comparison only.
   \item ``Near'' sample: $N_{\mathrm{sat}} = 24$\ satellite galaxies are drawn from a cone with the apex at a distance of 780\,kpc \cite[the distance at which we observe M31,][]{Conn2012}. The cone has an opening angle such that a circle of 150\,kpc radius is covered at the position of the host (approximately resembling the main area of the PAndAS survey). No satellites closer than $2.5^{\circ}$\ from M31 as seen from the artificial MW position are considered, in line with the procedure by \citet{Ibata2013}. We apply a radial cut such that only galaxies within a sphere of $d_{\mathrm{edge}}/2 = 426$\,kpc around the host can be selected. This sample is going to be compared to the GPoA parameters where Andromeda XXVII is assumed to be close to M31. We remind the reader that only 24 instead of 27 satellites are considered because the PAndAS survey area around M33, which contains two satellites in the observed sample, is not covered by the simulations and because Andromeda XVIII is too distant from M31 to fall into the spherical volume.
   \item ``Far'' sample: The third sample is similar to the second, but here only $N_{\mathrm{sat}} = 23$\ satellite galaxies are dawn. This assumes that Andromeda XXVII is far from M31, such that it is outside of the spherical volume covered by the simulated satellites. Consequently, when comparing with the observational data only planes consisting of $N_{\mathrm{GPoA}} = 14$\ satellites are searched in this sample. 
   \item ``Random'' sample: The last sample is identical to the third, but instead of drawing satellite positions and velocities from the simulation we first randomize both their position and their velocity direction by rotating them into a random direction. This results in them having a isotropic distribution around their hosts, but the same distribution of radial distances. Comparing the results for this and the previous two samples therefore allows us to determine the influence of the anisotropy present in the $\Lambda$CDM simulation on the results.
\end{enumerate}

Note that all four samples use the same viewing direction for each halo, enabling maximal comparability.
If the number of satellite galaxies found for one of the samples is too small, a new viewing direction is chosen for the halo. This step is repeated up to ten times. If by then no viewing  direction produced a sufficient number of satellites to lie within the artificial survey areas the halo is discarded from the analysis.
For each halo 10 different samples for each sample prescription are chosen, utilizing different random viewing directions. Applying all before-mentioned cuts, we arrive at a total of 15000 samples of satellite systems.

Each sample is then searched for the thinnest plane consisting of $N_{\mathrm{GPoA}} = 15$\ or 14 satellite galaxies. We follow a method similar to the one used by \citet{Conn2013} on the M31 satellite galaxy distribution. We determine the root-mean-square height of the $N_{\mathrm{GPoA}}$\ closest satellite galaxies to 10000 planes with normal vectors that are approximately evenly distributed on the sphere, while \citet{Conn2013} and BB14 (private communication) have used random plane orientations. The galaxy sample producing the smallest rms height is identified as the thinnest plane, and a full plane fit to its $N_{\mathrm{GPoA}}$\ members is performed using the method described in \citet{Pawlowski2013a}. In short, the principal axes of the satellite system are determined by calculating the eigenvectors of the tensor of inertia for the distribution of satellite galaxy positions. The rms of the height $r_{\mathrm{per}}$\ of the $N_{\mathrm{GPoA}}$\ satellites from the best-fitting plane and their rms distance $r_{\mathrm{par}}$\ after being projected into the best-fitting plane are determined. In addition, we measure the axis ratios $c/a$\ and $b/a$\ of short-to-long and intermediate-to-long axis. These quantities allow to discriminate between elongated prolate (both ratios are small) and plane-like oblate ($c/a$ is small but $b/a$ large) satellite distributions, in contrast to only using the parameters $r_{\mathrm{per}}$\ and $r_{\mathrm{par}}$. 

In addition, we determine the number of co-orbiting satellites within the plane sample by projecting the angular momenta of the satellite galaxies on to one normal vector of the best-plane fit. The co-orbiting number $N_{\mathrm{coorb}}$\ is then either the number of positive or negative angular momenta, whichever is larger. Note that according to this definition two satellites on very different (e.g. perpendicular) orbits are still considered co-orbiting if they both have angular momentum directions closer than $90^{\circ}$\ from the plane normal vector. The method therefore does not require the GPoA to be rotationally stabilized (which is the case if the angular momentum axes are aligned such that the structure is long-lived). If measurements of the proper motions of the M31 satellites reveal that the GPoA is indeed rotationally stabilized, such that the angular momentum directions cluster close to the plane normal direction (as in the case of the VPOS), then the significance of the satellite structure would be greatly increased.

The parameters determined for each sample of simulated satellites are then compared to those determined for that observed M31 satellite sample which fulfils the same selection criteria. The parameters of the observed satellite distribution have been determined using the same routine applied to the simulated samples, in particular the same cut on the distance from M31 was applied to the observed satellite sample. We use the most-likely distance moduli as homogeneously measured for all M31 satellites in the PAndAS field by means of the TRGB method \citep{Conn2012}. The resulting parameters are compiled in Table~\ref{tab:GPoAparams}.

For a consistent comparison, we have applied the same selection criteria to the observed positions of the M31 satellite galaxies. Out of the 27 satellites in the PAndAS sample used in the analyses of \citet{Ibata2013} and \citet{Conn2013}, two are outside of the main survey area with 150\,kpc radius at M31's distance. In addition, Andromeda XVIII is at a distance of $\approx 460$\,kpc from M31 and therefore outside of the volume considered around each of the model host galaxies. It consequently has to be removed from the sample as well. As discussed before, we furthermore discriminate between a near and a far position of Andromeda XXVII, which places it either inside or outside of the considered volume. Therefore we arrive at two sets of parameters, one assuming the distance modulus \citet{Richardson2011} report for Andromeda XXVII (lower limit) and one assuming the distance modulus of \citet{Conn2012} (upper limit).

\subsection{Results}

\begin{table}
  \caption{Fractions of simulated satellite systems reproducing the observed GPoA parameters.}
  \label{tab:GPoAresults}
  \begin{center}  
  \begin{tabular}{rlrrr}
  \hline
  \# & Criterion & $P_{\mathrm{sim}}^{\mathrm{near}}$ [\%] & $P_{\mathrm{sim}}^{\mathrm{far}}$ [\%] & $P_{\mathrm{rand}}^{\mathrm{far}}$ [\%] \\
 \hline
1 & $r_{\mathrm{per}}^{\mathrm{sim}} \leq r_{\mathrm{per}}^{\mathrm{obs}}$ & 10.55 & 15.85 & 3.80 \\
2 & $r_{\mathrm{par}}^{\mathrm{sim}} \geq r_{\mathrm{par}}^{\mathrm{obs}}$ & 25.58 & 23.26 & 21.43 \\
3 & $N_{\mathrm{coorb}}^{\mathrm{sim}} \geq N_{\mathrm{coorb}}^{\mathrm{obs}}$ & 4.77 & 1.79 & 0.28 \\
4 & $(c/a)^{\mathrm{sim}} \leq (c/a)^{\mathrm{obs}}$ & 10.73 & 14.03 & 4.11 \\
5 & $(b/a)^{\mathrm{sim}} \geq (b/a)^{\mathrm{obs}}$ & 17.28 & 25.59 & 30.29 \\
\hline
6 & Criteria 1 \& 2 & 1.49 & 2.05 & 0.26 \\
7 & Criteria 1 \& 3 & 0.90 & 0.59 & 0.01 \\
8 & Criteria 4 \& 5 & 0.31 & 0.77 & 0.15 \\
\hline
\it 9 & \it Criteria 1, 2 \& 3 & \it 0.17 & \it 0.09 & \it $< 0.007$ \\
\it 10 & \it Criteria 3, 4 \& 5 & \it 0.04 & \it 0.04 & \it $< 0.007$ \\
\hline
  & $N_{\mathrm{lopside}}^{\mathrm{sim}} \geq N_{\mathrm{lopside}}^{\mathrm{obs}}$ & 3.77 & 1.60 & 0.23 \\
  & $\alpha_{\mathrm{MW}}^{\mathrm{sim}} \leq \alpha_{\mathrm{MW}}^{\mathrm{obs}}$ & 6.83 & 7.23 & 6.89 \\ 
\hline
\end{tabular}
 \end{center}
 \small \medskip
 Fractions of simulated satellite systems fulfilling the criteria numbered in the first and specified in the second column. $P_{\mathrm{sim}}^{\mathrm{near}}$\ is the fraction of simulated systems fulfilling the criteria for the sample assuming Andromeda XXVII to be close to M31 (compared to the parameters given in column three of Table \ref{tab:GPoAparams}), $P_{\mathrm{sim}}^{\mathrm{far}}$\ those for the simulated systems selected assuming Andromeda XXVII to be far (compared to the parameters in column four of Table \ref{tab:GPoAparams}). $P_{\mathrm{rand}}^{\mathrm{far}}$\ is based on the same selection criteria as $P_{\mathrm{sim}}^{\mathrm{far}}$, but uses the randomized satellite positions and velocities. Only the criteria combinations printed in italics can be considered to fully reproduce a GPoA-like structure.
\end{table}

\begin{figure}
   \centering
   \includegraphics[width=84mm]{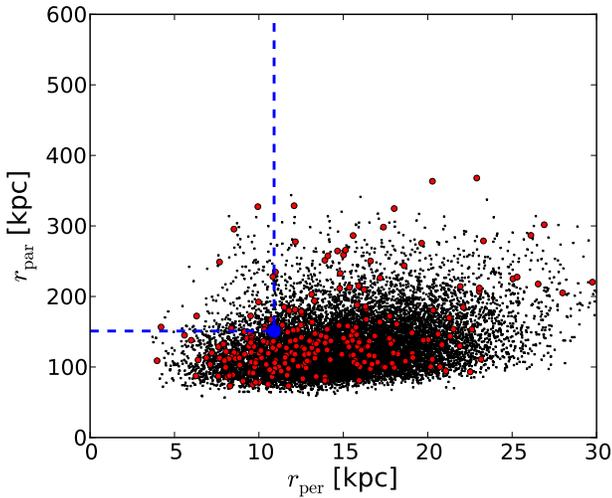}
   \caption{rms distance ($r_{\mathrm{par}}$) of the satellites projected on to their best-fitting plane plotted against the rms height ($r_{\mathrm{per}}$) of the satellites from the plane for the ``far'' sample. Satellite systems with a co-orbiting number $N_{\mathrm{coorb}}^{\mathrm{sim}} \geq N_{\mathrm{coorb}}^{\mathrm{obs}}$\ are plotted as red dots, all others as black points.
   The blue dot denotes the properties of the observed GPoA (determined for the satellite system after being subject to the same selection criteria as the simulated ones, see Table \ref{tab:GPoAparams} for the parameters). Only satellite systems falling in the area to the left and above the blue dashed lines reproduce the observed situation, and there are only 14 out of 15000 simulated systems (0.09 per cent, criterion \#9 in Table \ref{tab:GPoAresults}) which also contains a sufficient number of co-orbiting satellites.
 Compare to fig. 2 in BB14.   
   }
              \label{fig:bahl:rparrper}
\end{figure}

\begin{figure}
   \centering
   \includegraphics[width=70mm]{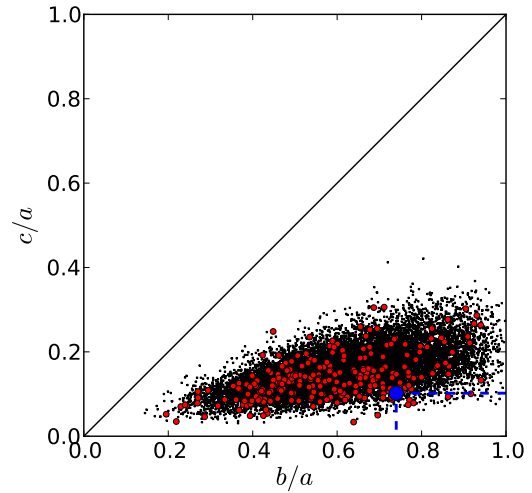}
   \caption{Axis-ratios of the GPoA-like planes in the simulates satellite systems for the ``far'' sample. 
   Satellite systems with a co-orbiting number $N_{\mathrm{coorb}}^{\mathrm{sim}} \geq N_{\mathrm{coorb}}^{\mathrm{obs}}$\ are plotted as red dots, all others as black points.
   The blue dot denotes the axis ratios of the observed GPoA (determined for the satellite system after being subject to the same selection criteria as the simulated ones, see Table \ref{tab:GPoAparams} for the parameters).   
   Only satellite systems falling into the area below and to the right of the blue dashed lines reproduce the observed situation, and there are only six out of 15000 simulated systems (0.04 per cent, criterion \#10 in Table \ref{tab:GPoAresults}) which also contains a sufficient number $N_{\mathrm{coorb}}$\ of co-orbiting satellites. These systems are not the same as the ones fulfilling all three criteria shown in Fig. \ref{fig:bahl:rparrper}.
   }
              \label{fig:bahl:axratios}
\end{figure}

\begin{figure*}
   \centering
   \includegraphics[width=180mm]{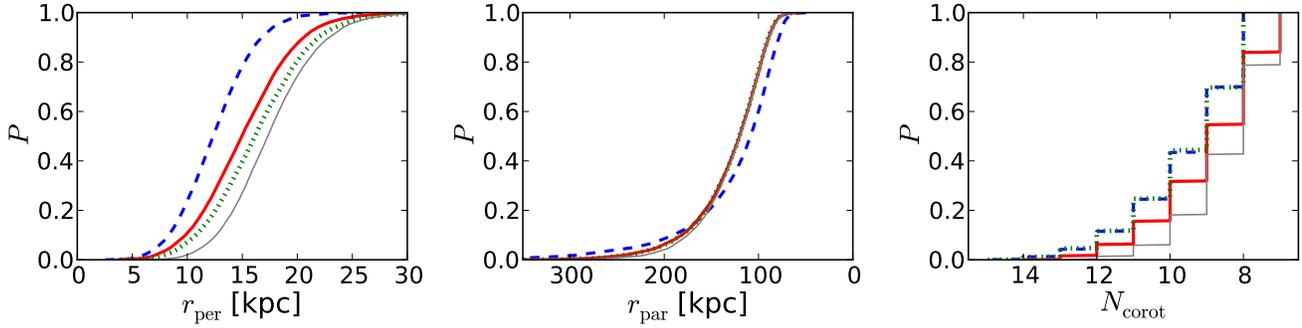}
   \caption{Cumulative probability functions of the rms height $r_{\mathrm{per}}$ of satellites from the best-fitting plane (left panel), their rms distance $r_{\mathrm{par}}$ when projected on to the plane (central panel) and the number of co-orbiting satellites $N_{\mathrm{coorb}}$ associated with the plane (right panel). Each panel shows four different distributions illustrating the effects of different sample selection prescription. Dashed blue line: $N_{\mathrm{sat}} = 27$, $N_{\mathrm{GPoA}} = 15$, artificial PAndAS area is modelled as a cylinder of radius 128\,kpc and satellites are drawn from a randomly oriented cube with edge length of $300/h$\,kpc. Solid red line: $N_{\mathrm{sat}} = 23$, $N_{\mathrm{GPoA}} = 14$. Dotted green line: $N_{\mathrm{sat}} = 24$, $N_{\mathrm{GPoA}} = 15$. Thin solid grey line: $N_{\mathrm{sat}} = 23$, $N_{\mathrm{GPoA}} = 14$, but drawing satellites not from the simulation directly but first randomizing their positions and velocities. In the latter three cases the artificial PAndAS survey is modelled as a cone with a radius of 150\,kpc at the distance of M31 and the satellites are drawn from a randomly oriented sphere with radius $300/h$\,kpc.
   It is more likely to find narrow planes (small $r_{\mathrm{per}}$) when more satellite galaxies are drawn from a more narrow volume.
   Drawing sample galaxies from a cube produces more satellite systems with large $r_{\mathrm{par}}$\ values, as sometimes the random line of sight is oriented along two opposing corners.
   In satellite planes consisting of 14 instead of 15 objects the probability of finding large numbers of co-orbiting satellites is lower.
   }
              \label{fig:bahl:comparison}
\end{figure*}

The results of our analysis are compiled in Table \ref{tab:GPoAresults}. The table includes only the last three sample prescriptions, the first one is incompatible with the observed satellite distribution and can therefore not be compared with it. The second column of the table lists the criteria according to which a simulated satellite system is considered to be similar to the observed one, which are numbered in the first column. The five criteria are :
\beq
r_{\mathrm{per}}^{\mathrm{sim}} \leq r_{\mathrm{per}}^{\mathrm{obs}},
\eeq
\beq 
r_{\mathrm{par}}^{\mathrm{sim}} \geq r_{\mathrm{par}}^{\mathrm{obs}},
\eeq
\beq 
N_{\mathrm{coorb}}^{\mathrm{sim}} \geq N_{\mathrm{coorb}}^{\mathrm{obs}},
\eeq
\beq 
(c/a)^{\mathrm{sim}} \leq (c/a)^{\mathrm{obs}},
\eeq
\beq  
(b/a)^{\mathrm{sim}} \geq (b/a)^{\mathrm{obs}}. 
\eeq
The first two criteria are quite similar to the last two because they measure either the absolute or the relative extend of the structure, so it does not make too much sense to combine the four of them. But either of these two combinations (1-2 or 4-5) should be combined with the third criterion, since all three criteria were essential in the discovery of the feature by \citet{Ibata2013}. Note again that BB14 have only considered combinations of two of these criteria without providing a motivation for this choice. The combined criteria (numbered 6 to 10 in Table~\ref{tab:GPoAresults}) then have to be met simultaneously by a satellite sample. The last two columns in Table \ref{tab:GPoAresults} give the fractions of systems that contain GPoA-like planes which fulfil the respective criteria. The systems consist of either the 24 brightest satellites from the simulation that lie within the selection volume of which 15 are considered for the plane fit ($P_{\mathrm{sim}}^{\mathrm{near}}$) or the 23 brightest satellites from the simulation that lie within the selection volume of which 14 are considered for the plane fit ($P_{\mathrm{sim}}^{\mathrm{far}}$). 

In the following we focus on the ``far'' sample prescription, which assumes the same distance ($\approx 1250$\,kpc) to Andromeda XXVII as \citet{Ibata2013} and \citet{Conn2013}, but we note that the results for the sample prescription assuming a closer distance to that satellite are similar (see the results compiled in Table \ref{tab:GPoAresults}).

\subsubsection{Comparing the plane thickness, radial extend and co-orbiting sense}

Fig. \ref{fig:bahl:rparrper} plots the rms distance $r_{\mathrm{par}}$ of the satellites associated with the planes versus the rms thickness $r_{\mathrm{per}}$ of the fitted planes similar to fig. 2 in BB14. In contrast to these authors we cannot find $r_{\mathrm{par}}$\ values much larger than 300\,kpc because our method only considers satellites within 426\,kpc from the host avoiding satellites with larger distances that are situated in the corners of the cube extracted from the simulation.

To be comparable to the observed GPoA, a simulated satellite population must contain a plane of satellites which is similarly narrow. Like BB14 we first measure this using the rms height $r_{\mathrm{per}}$\ of the fitted satellite plane, i.e. $r_{\mathrm{per}}^{\mathrm{sim}} \leq r_{\mathrm{per}}^{\mathrm{obs}}$. 
Only 15.9 per cent of the simulated satellite samples have this property. That number is much lower than 40 per cent reported by BB14, even though we are using the same data. The difference indicates that our adjustments to the selection of simulated satellites, i.e. applying the same selection to the observed satellite sample and the simulated satellites, have a significant influence on the result. Only 3.8 per cent of the sample using the randomized satellite positions fulfil this criterion, indicating that the simulated satellite systems are not fully isotropic.
This is to be expected, because dark matter haloes and their sub-halo systems are not perfectly spherical, due to their merger history and the preferential infall of mass along dark matter filaments \citep[e.g.][]{Libeskind2011}. That some non-isotropic satellite distributions have a larger likelihood to form satellite planes has already been demonstrated by \citet{Metz2007}. However, the non-isotropy of the simulated satellite systems is generally insufficient to explain the strong observed coherence of the satellite galaxy positions and velocities \citep[e.g.][]{Pawlowski2012b}.

We find $r_{\mathrm{par}}^{\mathrm{sim}}$ is larger than $r_{\mathrm{par}}^{\mathrm{obs}}$ in 23.3 per cent of all simulated satellite samples. This fraction is again significantly different from the 10 per cent reported by BB14. The reason for the discrepancy is that we determine $r_{\mathrm{par}}$\ only for those satellites in the GPoA that are close enough to M31 to be within the considered simulation volumes. This either excludes Andromeda XXVII or assumes it to be close to M31. Both scenarios result in a considerably smaller $r_{\mathrm{par}}^{\mathrm{obs}}$\ than the one apparently used by BB14 \footnote{BB14 do not mention the adopted value, but looking at their fig. 2 it appears to be approximately 190\,kpc.}. This second criterion is fulfilled by a similar fraction of 21.4 per cent of the samples using randomized satellite positions, as to be expected because these satellite systems have the same radial distribution.

To resemble the planar GPoA, criteria 1 and 2 have to be met simultaneously. Only 2.0 per cent of all simulated satellite systems do so (criterion 6 in Table \ref{tab:GPoAresults}), a value consistent with the 2 per cent reported by BB14. In addition, we know that the GPoA satellites preferentially co-orbit. The number of co-orbiting satellites in the plane, $N_{\mathrm{coorb}}$, must be at least as high as observed: $N_{\mathrm{coorb}}^{\mathrm{sim}} \geq N_{\mathrm{coorb}}^{\mathrm{obs}}$. This individual criterion (number 3) is met by only 1.8 per cent of the simulated satellite samples. Only 0.28 per cent of the randomized satellite systems fulfil this third criterion.

Only when the third criterion ($N_{\mathrm{coorb}}$) is combined with the other two ($r_{\mathrm{per}}$ and $r_{\mathrm{par}}$) can a simulated satellite system be considered similar to the GPoA (although this is still ignoring the alignment of the GPoA with the line-of-sight and the lopsidedness of the observed satellite distribution). Unfortunately BB14 do not discuss this criteria combination. We find that only 0.09 per cent (14 out of 15,000 samples) meet all three criteria simultaneously (criterion 9 in Table \ref{tab:GPoAresults}). 

Finally, none of the 15,000 systems based on the randomized satellite positions reproduce all three criteria. From this we can estimate an upper limit of 0.007 per cent, which is consistent with \citet{Ibata2013}, who report a significance of 99.998 per cent for the GPoA, meaning that a fully isotropic satellite distribution can reproduce the observed properties in only 0.002 per cent of all cases.

\subsubsection{Comparing the plane axis ratios and co-orbiting sense}

The axis ratios of the fitted planes are plotted in Fig. \ref{fig:bahl:axratios}. Most simulated satellite systems have a ratio of short-to-long axis $(c/a)$\ in the range of 0.1 to 0.3, but they show a wide spread in the intermediate-to-long axis ratio $b/a$. Most importantly, there is a distinct trend of larger $c/a$\ for larger $b/a$. This indicates that narrow distributions are more prolate while more oblate (and therefore plane-like) distributions are wider along their shortest axis. The danger of misinterpreting elongated, prolate satellite distributions as planes similar to the GPoA when only looking at the rms height is therefore real.

For a simulated satellite population to display a similarly narrow alignment as the observed GPoA, $(c/a)$\ has to be at least as small as the observed value, i.e.  $(c/a)^{\mathrm{sim}} \leq (c/a)^{\mathrm{obs}}$. In addition, for the alignment to be similarly plane-like, $b/a$\ has to be at least as large as for the observed system, i.e. $(b/a)^{\mathrm{sim}} \geq (b/a)^{\mathrm{obs}}$. Only 0.8 per cent of the simulated satellite samples fulfil these two criteria simultaneously. In addition, to resemble the co-orbiting satellite plane the $N_{\mathrm{coorb}}$\ criterion has to be fulfilled too. Only six, i.e. 0.04 per cent, of all simulated satellite systems fulfil these three criteria simultaneously and can be considered to resemble the observed GPoA (not accounting for the alignment with the line-of-sight and the lopsidedness). This number is lower than the one found for the previously discussed three-criteria combination and corresponds to different satellite systems.
Again, none of the randomized satellite systems fulfil all three criteria simultaneously.

\subsubsection{Effects of sample selection}

Fig. \ref{fig:bahl:comparison} compares the cumulative probability distributions of the three quantities $r_{\mathrm{per}}$, $r_{\mathrm{par}}$\ and $N_{\mathrm{coorb}}$ for the four samples as defined in Sect. \ref{sect:bahl:method}. The major difference is that for the sample prescriptions constructed for consistency with the observed satellite system (near and far samples, ii and iii in Sect. \ref{sect:bahl:method}), the rms height of the planes are considerably larger than for the sample selection used by BB14. This is because the latter draws a larger number of satellites than observed from a survey volume which is smaller than the observed one. 
The sample using randomized satellite positions results in even larger $r_{\mathrm{per}}$.
Thus, the average rms heights of the fitted planes depend strongly on the sample selection recipe. It is smallest for the BB14-like sample ($\langle r_{\mathrm{per}} \rangle_{\mathrm{BB14}} = 12.46$\,kpc) and larger for the samples containing fewer satellites ($\langle r_{\mathrm{per}} \rangle_{\mathrm{near}} = 16.33$\ and $\langle r_{\mathrm{per}} \rangle_{\mathrm{far}} = 15.15$\,kpc). The difference between $\langle r_{\mathrm{per}} \rangle_{\mathrm{BB14}}$\ and $\langle r_{\mathrm{per}} \rangle_{\mathrm{far}}$\ is similar to the difference between the latter and $\langle r_{\mathrm{per}} \rangle_{\mathrm{rand}} = 17.57$. The average axis ratios show the same behaviour.

In addition, for the sample prescriptions constructed to be consistent with the observed satellite sample, the probability of finding large values of $r_{\mathrm{par}}$\ is lower while that of finding small values is increased. This clearly shows the effect of choosing satellites either from the randomly oriented cube or from a sphere embedded in them. However, the average rms distance of the GPoA satellites projected into the plane does not depend strongly on the sample selection, the averages are similar for all samples: $\langle r_{\mathrm{par}} \rangle_{\mathrm{BB14}} = 125$\,kpc , $\langle r_{\mathrm{par}} \rangle_{\mathrm{near}} = 129$\,kpc , $\langle r_{\mathrm{par}} \rangle_{\mathrm{far}} = 129$\,kpc and $\langle r_{\mathrm{par}} \rangle_{\mathrm{rand}} = 127$\,kpc.

As expected the number of co-orbiting satellites tends to be lower for the sample considering only $N_{\mathrm{GPoA}} = 14$\ satellites to belong to the GPoA instead of 15. It is also much lower for the sample using randomized satellite positions and velocity directions. The average number of co-orbiting satellites in the GPoA-like planes depends mostly on the number of satellites in the GPoA. The samples using 15 GPoA members result have $\langle N_{\mathrm{coorb}} \rangle_{\mathrm{BB14}} = 9.56$\ and $\langle N_{\mathrm{coorb}} \rangle_{\mathrm{near}} = 9.57$, while the one using only 14 satellites for the plane fit has $\langle N_{\mathrm{coorb}} \rangle_{\mathrm{far}} = 8.95$. Randomizing the satellite positions and velocity directions reduces this number by about 0.5, to $\langle N_{\mathrm{coorb}} \rangle_{\mathrm{rand}} = 8.48$.

\subsubsection{Additional properties}

In addition to the previously-discussed parameters, the GPoA exhibits two more puzzling properties. It is seen almost edge-on (the angle $\alpha_{\mathrm{MW}}$\ between the line-of-sight and the GPoA is small) and the distribution of satellites belonging to the GPoA is lopsided, i.e. a significantly larger number (13 of the 15 in the GPoA) of satellite galaxies ($N_{\mathrm{lopside}}$) is on the near side of M31. Among the simulated satellite samples, 1.6 per cent show a similar degree of lopsidedness ($N_{\mathrm{lopside}}^{\mathrm{sim}} \geq N_{\mathrm{lopside}}^{\mathrm{obs}}$) and 7.2 per cent align similarly well with the MW ($\alpha_{\mathrm{MW}}^{\mathrm{sim}} \leq \alpha_{\mathrm{MW}}^{\mathrm{obs}}$). In combination, these two criteria are met by 0.13 per cent of the samples and combining them with either of the the previous three-criteria combinations results in no sample fulfilling the five criteria simultaneously. However, we note that these two features might be related to the dynamics of the LG and the MW's particular viewing position which are not considered in the selection of host galaxies and their viewing directions.

\subsection{Summary}

A satellite galaxy configuration as observed in the GPoA around M31, satisfying simultaneously the three criteria defined by BB14, is only found in 0.04 to 0.17 per cent of the satellite samples based on the Millennium II simulation, depending on Andromeda XXVII's distance to M31. Additionally to the more realistic sample selection, the main reason for this difference of our results from those of BB14 is that we require GPoA-like planes to agree with all three observed properties, instead of arbitrarily combining only two of them.

Our result is robust against changing the prescription of the major GPoA properties. Comparing the rms height $r_{\mathrm{per}}$, rms distance of the satellites projected into the plane $r_{\mathrm{per}}$\ and their co-orbiting number $N_{\mathrm{coorb}}$\ yields similarly low probabilities as a comparison of the axis ratios $c/a$\ and $b/a$\ in combination with $N_{\mathrm{coorb}}$.

Further improvements of the analysis are expected from (i) using higher-resolution simulations so the analysis does not have to rely on orphan galaxies and (ii) a comparison of the observed satellite distribution of large galaxies in LG-like galaxy groups. These points will be addressed in a future paper.

While our work was being refereed, \citet{Ibata2014} have also published a re-analysis of the work by BB14. While our analysis stays as close as possible to the one performed by BB14, the analysis of \citet{Ibata2014} remains as close as possible to the one performed on the observational data by \citet{Ibata2013}. They apply the exact PAndAS survey area to the simulations, enforce the satellite planes to be seen edge-on as is the case for observed M31 system and in addition check whether the specific angular momenta of the simulated satellites in the GPoA-like planes are as large as observed. It is reassuring that the results of both our studies with different approaches agree: \citet{Ibata2014} find that only 0.04 per cent of the planes in satellite systems drawn from the Millennium II simulation can reproduce the thickness, radial extend and velocity coherence of the GPoA, to be compared with our result of 0.04 to 0.17 per cent.

%__________________________________________________________________

\section{Searching for a MW VPOS-like structure in $\Lambda$CDM simulations}
\label{sect:wang}

\subsection{Previous method and its limitations}
\label{sect:wang:problems}

W13 report that 6--13 per cent of satellite systems in cosmological simulations are as flat as that of the MW VPOS. However, there are a number of indications that this over-estimates the probability of VPOS-like structures to appear in $\Lambda$CDM simulations. The major points and our adjustments to the analysis are given below.
\begin{enumerate}
  \item Arguably the most important property of the observed MW satellite system is that a significant fraction of the satellites co-orbit within the VPOS. This orbital alignment indicates that it is rotationally stabilized and therefore 
  could be long-lived \citep{LyndenBell1976,Metz2008,Pawlowski2012a,Pawlowski2013b}. W13 have not tested whether the simulated satellite systems show a similar degree of phase-space correlation. Because the velocities of satellite galaxies in simulations are known, we have to closely pay attention to this observational constraint on the VPOS properties and will therefore incorporate it in the following analysis.
  \item W13 analyse satellite systems in the large-scale Millennium II simulation as well as those around the six main haloes in the high-resolution Aquarius simulations \citep{Springel2008}. None of the Aquarius satellite systems they analyse can reproduce the same amount of flattening as measured for the VPOS using either $c/a$\ (their fig. 7) or $r_{\mathrm{per}}$\ (their fig. 10). As noted by W13, their conclusion that 6--13 per cent of the satellite systems can be as flattened as the VPOS is only based on satellite systems constructed by applying a semi-analytic galaxy formation model to the Millennium II simulation \citep{Guo2011}. This might indicate that the inclusion of orphan galaxies with incorrect positions affects the analysis, but the reason could also be small number statistics as there are only six host galaxies in the Aquarius simulation suite. In addition, W13 use satellites obtained from the dark matter-only simulations using slightly different semi-analytical models for the Aquarius \citep{Cooper2010} and Millennium II \citep{Guo2011} simulations, while in addition using an abundance matching technique for the Aquarius simulations, too. In the following we will focus on the Millennium II simulation.
  \item  W13 compare the observed and simulated satellite systems using two different measures of the flattening: the axis ratios of the satellite distribution $c/a$, where the observed value is met by 6 per cent of their simulated systems, and the ratio of the rms thickness $r_{\mathrm{per}}$\ of the best-fitting plane to a fixed cut-off radius of $r_{\mathrm{cut}} = 250$\,kpc. As already pointed out in \citet{Pawlowski2013a}, the latter method, according to which 13 per cent of the simulated systems are meeting the observed value, is biased. It does not consider the radial distribution of the satellites, and thus a radially more concentrated simulated satellite system could count as VPOS-like even though the satellites are distributed nearly isotropically. We will account for this by using the same method as in Sect. \ref{sect:bahl} and demanding the satellites to be sufficiently extended radially, as measured by $r_{\mathrm{par}}$.
  \item Another important aspect influencing the result of W13 is how they incorporate the obscuration by the Galactic disc in their satellite selection. They assume that no satellite galaxy can be observed within a critical angular distance from the MW disc ($\theta_{\mathrm{crit}} = 9.5^{\circ}$), and model this selection effect by placing obscuring discs into their simulated satellite systems, ignoring all satellites within $\theta_{\mathrm{crit}}$\ of them. Instead of using randomly oriented obscuring discs they iteratively orient them such that they are perpendicular to the major axis of the satellite distribution. While the observed VPOS is indeed almost perpendicular to the MW plane, there is no reason to expect such an orientation a priori. 
In fact, observations and cosmological simulations suggest the contrary, i.e. a preferred planar alignment of satellites with the major axis of their host galaxies \citep[e.g.][]{Sales2004,Sales2009,Brainerd2005,Bett2010,Lovell2011}.
 Iteratively forcing the obscuring disc to be perpendicular to the major axis of the satellite distribution has the effect of obscuring those satellite galaxies which are perpendicular (and therefore on average most-distant) from the best-fitting satellite plane. This artificially increases the measured flattening of the system, biasing them to find an agreement with the observed, narrow VPOS. 
 If the reason for the VPOS is the absence of satellites in the disc of the MW (Zone of Avoidance), then incorporating randomly oriented obscuring discs already accounts for the effect of the MW disc on the satellite distribution and should result in finding plenty of VPOS-like structures in the simulated satellite systems.
  \item When determining the properties of the observed MW satellite system which will be compared with those of the simulated systems, W13 decide to use the 11 brightest, i.e. classical MW satellite galaxies. They note that the 12th-brightest MW satellite Canes Venatici I (CVnI) has only a slightly lower luminosity than the next-brighter classical satellites Draco and Ursa Minor ($M_{\mathrm{V}} = -8.6$\ compared to $-8.8$, \citealt{McConnachie2012}), but argue that the inclusion of CVnI in the MW satellite sample has almost no effect on the axis ratio of the satellite system and do not include it in their analysis. However, one of their results is that simulated satellite systems become less flattened when more satellites are considered (see e.g. their fig. 9). Thus, even though the flattening might be the same for 11 or 12 \textit{observed} satellites, there might be a difference between the flattening of 11 and 12 \textit{simulated} satellites. Whether the addition of one object changes the results might be doubted, but should be tested. We will do so in the following.
\end{enumerate}

Following the same procedure as in the previous section, we account for these issues with a refined analysis method and compare the results to those which W13 obtained using their method.

\subsection{New method}
\label{sect:wang:method}

\begin{table*}
 \begin{minipage}{170mm}
  \caption{Parameters of the observed VPOS}
  \label{tab:VPOSparams}
  \begin{center}
  \begin{tabular}{llrrr}
  \hline
  Quantity & Description & Classical Sats. & Classical Sats. + CVnI\\
 \hline
$N_{\mathrm{VPOS}}$ & Number of satellites used for the VPOS plane fit & 11 & 12 \\
$r_{\mathrm{per}}$ & rms height of satellites from best-fitting plane & 19.6\,kpc & 20.8\,kpc \\
$r_{\mathrm{par}}$ & rms distance of satellites projected into best-fitting plane & 129.5\,kpc & 138.9\,kpc \\
$c/a$ & Minor to major axis rms height ratio & 0.182 & 0.183 \\
$b/a$ & Intermediate to major axis rms height ratio & 0.508 & 0.530 \\
$\Delta_{\mathrm{std}}$ & spherical standard distance of 8 most-concentrated orbital poles & $29.3^{\circ}$ & $\leq 29.3^{\circ}$ \\
$\theta_{\mathrm{VPOS}}$ & Angle between VPOS normal and average direction of 8 most-concentrated orbital poles & $18.9^{\circ}$ & $24.5^{\circ}$ \\
\hline
\end{tabular}
 \end{center}
 \small \medskip
  The VPOS plane parameters using the observed positions and line-of-sight velocities of the brightest MW satellite galaxies using the data compiled by \citet{McConnachie2012}. They have been determined using the same method applied to the simulated satellite systems. The first column gives the symbol of the quantity described in the second column. The third column gives the parameters of the plane fitted to the 11 classical satellite galaxies, while the fourth column gives those of the plane fitted to these satellites plus the next-fainter satellite CVnI.
\end{minipage}
\end{table*}

For the following analysis we make use of the same data from the Millennium II simulation as in Sect. \ref{sect:bahl}. This results in a similar main halo selection as used by W13, but in a slightly smaller range in virial mass of 1.1 to $1.7 \times 10^{12}\,\mathrm{M}_{\sun}$ instead of 1.0 to $2.0 \times 10^{12}\,\mathrm{M}_{\sun}$.  Working with the same main halo sample has the advantage of a direct comparison of the probabilities for finding either a VPOS or a GPoA-like system among the same main galaxy population. This makes it possible to decide which of the structures is the least common in the $\Lambda$CDM satellite systems we analyse. By combining the probabilities of finding a VPOS or a GPOA we are also able to determine the probability to find both structures in the LG, under the reasonable assumption that both the MW and M31 have halo masses within the same range.

We only consider satellites with distances from their host in the range of 15 to 260\,kpc. The upper limit was chosen to be compatible with the Galactocentric distance of Leo I, the most distant of the 11 classical satellite galaxies. For each main halo, we select the 11 satellite galaxies with the largest baryonic masses and with an angular separation $\theta>\theta_{\mathrm{crit}}$ from a randomly oriented plane, imitating the obscuration of the Galactic disc. We chose the same obscuration angle as W13, $\theta_{\mathrm{crit}} = 9.5^{\circ}$. However, in contrast to their analysis we do not require that this galactic plane is perpendicular to the minor axis of the plane fitted to the satellite galaxies, for reasons we discussed before.

In total there are 1693 main haloes in the chosen mass range which contain a sufficient number of satellite galaxies. As in Sect. \ref{sect:bahl} we generate 10 samples for each main halo, by adopting different random obscuring disc orientations, such that we have 16930 simulation samples.

For each sample we determine the best-fitting plane of all satellites using the method described in Sect. \ref{sect:bahl}. Again we compute the relevant properties (in particular $r_{\mathrm{per}}$, $r_{\mathrm{par}}$, $c/a$\ and $b/a$) and compare them to those derived for the observed MW satellite galaxy system. Further, we determine how strongly the satellites are co-orbiting employing the method presented in \citet{Pawlowski2012b}. For this purpose, we construct the orbital poles $\hat{n_i}$ (directions of angular momenta) of the simulated satellite galaxies and determine which of all possible combinations of eight of these poles for a given satellite system have the smallest spherical standard distance $\Delta_{\mathrm{std}}$\ \citep{Metz2007,Pawlowski2012b} from their average direction $\langle \hat{n} \rangle$,
$$
\Delta_{\mathrm{std}} = \sqrt{ \frac{ \sum_{i=1}^{8} \left[ \arccos \left( \langle \hat{n} \rangle \cdot \hat{n}_{i} \right) \right] ^{2} }{ 8 } } = \min.
$$
These eight orbital poles are the best-aligned (or best-concentrated) orbital poles of the sample, and we record their concentration as measured by the spherical standard distance $\Delta_{\mathrm{std}}$\ as well as the angle between their average direction and the normal vector of the best-fitting plane $\theta_{\mathrm{VPOS}}$. As in \citet{Pawlowski2013b} we checked that considering only the six best-aligned poles does not change the result.

In addition to the Millennium II data we also compare our results to randomized satellite systems. For these, we rotate both the positions and velocities of the satellite galaxies into random directions before applying the obscuration-cut. This process leads to isotropic satellite systems with the same radial distribution as in the non-randomized case.

To test the statistical robustness of our results we also select systems of 12 satellite galaxies to be compared to the properties of plane fitted to the 11 classical MW satellite galaxies plus CVnI, which has a very similar luminosity as the least-luminous classical satellites Draco and Ursa Minor (as discussed in \ref{sect:wang:problems}). Because CVnI was discovered in the SDSS survey at large Galactic latitude ($b = + 79.8^{\circ}$), we require the 12th satellite galaxy drawn from the simulations to be within $60^{\circ}$\ from the north pole direction of the artificial obscuring disc. 

Table \ref{tab:VPOSparams} compiles the properties of the VPOS planes fitted to the observed MW satellite galaxies for samples consisting of the 11 classical satellite galaxies and for the sample extended with CVnI.
The orbital pole of CVnI is not known, so we adopt the same orbital pole concentration and average direction as determined from the smaller sample of only 11 satellites. We will compare these properties with those of the larger simulation samples, i.e. we will compare the concentration of the eight best-aligned orbital poles out of only 11 observed satellite galaxies with the concentration of the eight best-aligned orbital poles out of 12 poles in the simulation. The benefit is that our results will remain valid once CVnI's proper motion is known, even in the unlikely case that it is not co-orbiting in the VPOS. Our results are therefore lower limits, because assuming it to orbit in VPOS, CVnI's predicted proper motion would make it align better than some of the known poles \citep{Pawlowski2013b}.

\subsection{Results}
\label{sect:wang:results}

\begin{table}
  \caption{Fractions of simulated satellite systems reproducing the observed VPOS parameters.}
  \label{tab:VPOSresults}
  \begin{center}
  \begin{tabular}{rlrrr}
  \hline
  \# & Criterion & $P_{\mathrm{sim11}}$ [\%] & $P_{\mathrm{rand11}}$ [\%] & $P_{\mathrm{sim12}}$ [\%] \\
 \hline
1 & $r_{\mathrm{per}}^{\mathrm{sim}} \leq r_{\mathrm{per}}^{\mathrm{obs}}$ & 4.45 & 2.10 & 4.44 \\
2 & $r_{\mathrm{par}}^{\mathrm{sim}} \geq r_{\mathrm{par}}^{\mathrm{obs}}$ & 51.65 & 49.69 & 34.90 \\
3 & $(c/a)^{\mathrm{sim}} \leq (c/a)^{\mathrm{obs}}$ & 3.35 & 0.92 & 2.39 \\
4 & $(b/a)^{\mathrm{sim}} \geq (b/a)^{\mathrm{obs}}$ & 86.36 & 92.86 & 84.05 \\
5 & $\Delta_{\mathrm{std}}^{\mathrm{sim}} \leq \Delta_{\mathrm{std}}^{\mathrm{obs}}$ & 2.48 & 0.16 & 4.11 \\
6 & $\theta_{\mathrm{VPOS}}^{\mathrm{sim}} \leq \theta_{\mathrm{VPOS}}^{\mathrm{obs}}$ & 13.44 & 10.06 & 24.11 \\
\hline
7 & Criteria 1 \& 2 & 1.22 & 0.27 & 0.63 \\
8 & Criteria 3 \& 4 & 1.57 & 0.64 & 1.02 \\
9 & Criteria 1 \& 5 & 0.35 & $< 0.006$ & 0.41 \\
10 & Criteria 3 \& 5 & 0.28 & 0.01 & 0.27 \\
11 & Criteria 5 \& 6 & 0.97 & 0.05 & 1.92 \\
\hline
12 & Criteria 1, 2 \& 5 & 0.15 & $< 0.006$ & 0.11 \\
13 & Criteria 3, 4 \& 5 & 0.12 & $< 0.006$ & 0.08 \\
14 & Criteria 1, 5 \& 6 & 0.15 & $< 0.006$ & 0.23 \\
15 & Criteria 3, 5 \& 6 & 0.12 & 0.01 & 0.15\\
\hline
\it 16 & \it Criteria 1, 2, 5 \& 6 & \it 0.06 & \it $< 0.006$ & \it 0.04 \\
\it 17 & \it Criteria 3, 4, 5 \& 6 & \it 0.05 & \it $< 0.006$ & \it 0.05 \\
\hline
\end{tabular}
 \end{center}
 \small \medskip
  Fractions of simulated satellite systems fulfilling the criteria numbered in the first column and specified in the second. $P_{\mathrm{sim11}}$\ gives the fraction of simulated systems fulfilling the criteria for the 11 brightest satellites not hidden by the obscuring MW disc, $P_{\mathrm{rand11}}$\ those for the 11 brightest satellites with randomized position and velocity directions (both compared to the parameters given in column three of Table \ref{tab:VPOSparams}). $P_{\mathrm{sim12}}$\ gives the fractions for the simulated satellite systems consisting of the 11 brightest satellites outside the obscured region and the next-fainter one within the approximate SDSS survey area. (i.e. compared to the parameters determined including CVnI given in column four of Table \ref{tab:VPOSparams}).
  Only the criteria combinations printed in italics can be considered to fully reproduce a VPOS-like structure.
\end{table}

The results of our analysis are compiled in Table \ref{tab:VPOSresults}. As in Sect. \ref{sect:bahl}, the second column of the table lists the criteria according to which a simulated satellite system is considered to be similar to the observed one, numbered in the first column. Combined criteria (numbers 7 to 17) have to be met simultaneously. The last three columns give the fractions of systems fulfilling the respective criteria which consist of either the 11 brightest satellites in the simulation outside of the obscuration region ($P_{\mathrm{sim11}}$), the 11 brightest outside the obscuration area for randomized angular positions and velocities of the satellites ($P_{\mathrm{rand11}}$) and for the 12 brightest satellites in the simulation and outside the obscured region, of which the least-luminous one lies within the approximate SDSS footprint ($P_{\mathrm{sim12}}$).

\subsubsection{A thin VPOS}

\begin{figure}
   \centering
   \includegraphics[width=84mm]{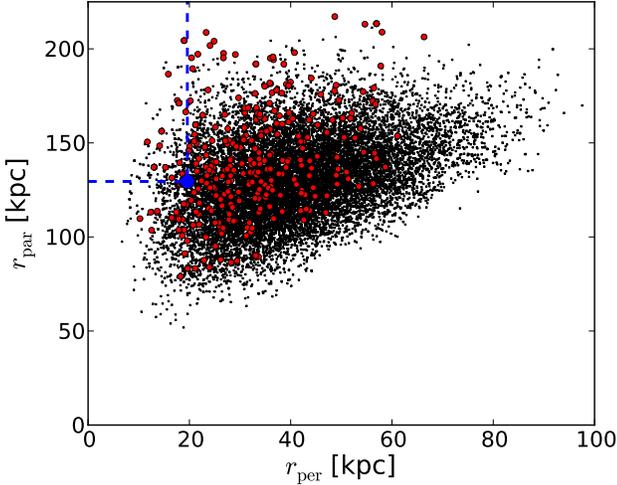}
   \caption{rms distance $r_{\mathrm{par}}$\ of the satellites projected into their best-fitting plane plotted against the rms height $r_{\mathrm{per}}$\ of the satellites from the plane for the samples of 11 satellites. Those satellite systems which have a sufficiently concentrated orbital pole distribution $\Delta_{\mathrm{std}}^{\mathrm{sim}} \leq \Delta_{\mathrm{std}}^{\mathrm{obs}}$\ are plotted as red dots, all others as black points.
   The blue dot denotes the axis ratios of the observed classical MW satellites. Only satellite systems falling into the area to the left and above of the blue dashed lines reproduce the observed spatial distribution. Only 1.22 per cent of the sample systems fall into this range (criterion \#7 in Table \ref{tab:VPOSresults}), and only 0.15 per cent simultaneously have sufficiently concentrated orbital poles (criterion \#12 in Table \ref{tab:VPOSresults}).
   }
              \label{fig:wang:rparrper}
\end{figure}

\begin{figure}
   \centering
   \includegraphics[width=70mm]{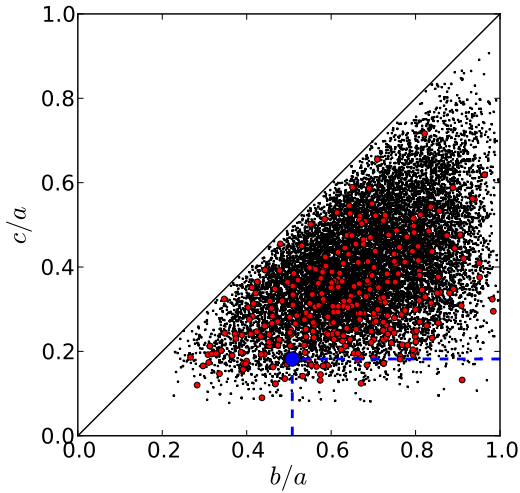}
   \caption{Axis-ratios of the planes fitted to the simulated satellite samples for the samples of 11 satellites. 
   Those satellite systems which have a sufficiently concentrated orbital pole distribution $\Delta_{\mathrm{std}}^{\mathrm{sim}} \leq \Delta_{\mathrm{std}}^{\mathrm{obs}}$\ are plotted as red dots, all others as black points.
   The blue dot denotes the axis ratios of the observed classical MW satellites. Only satellite systems falling into the area to the right and below by the blue dashed lines reproduce the observed spatial flattening. Only 1.57 per cent of the sample system fall into this range (criterion \#8 in Table \ref{tab:VPOSresults}) and only 0.12 per cent simultaneously have sufficiently concentrated orbital poles (criterion \#13 in Table \ref{tab:VPOSresults}). 
    Compare to fig. 8 in W13.   
   }
              \label{fig:wang:axratios}
\end{figure}

W13 have reported that 6 per cent of the simulated satellite systems have a smaller axis ratio $c/a$\ than observed (our criterion 3 in Table \ref{tab:VPOSresults}). For the observed MW satellites, they measure a value of $c/a = 0.18$, while we use $c/a = 0.182$. We therefore use the same criterion, but find only a fraction of 3.4 per cent of the samples to fulfil this criterion.
In addition to this relative shape of the distribution as measured by the axis ratio, W13 also test how often their satellite systems are as thin as the observed one by using the rms height $r_{\mathrm{per}}$\ of the satellites from the best-fitting plane as measured in kpc. They compare the ratio of $r_{\mathrm{per}}$\ to a fixed cut-off-radius of $R_{\mathrm{cut}} = 250$\,kpc, chosen to be similar to the distance of the most-distant known MW satellite Leo I. As $R_{\mathrm{cut}}$\ is kept at a fixed value independent of the simulated satellite systems, this method is identical to comparing the $r_{\mathrm{per}}$-values of the observed and the simulated satellite systems directly, corresponding to our criterion~1 in Table \ref{tab:VPOSresults}.
W13 give a value of $\frac{r_{\mathrm{per}}}{R_{\mathrm{cut}}} = 0.074$, which corresponds to $r_{\mathrm{per}} = 18.5$\,kpc, whereas we find $r_{\mathrm{per}} = 19.6$\,kpc. Our criterion is therefore slightly less strict.
In contrast to W13, who report 13 per cent of their systems are having a thickness similar to the observed one, we find this situation is encountered in only 4.5 per cent of our samples.
Part of these discrepancies will be due to the enforced polar alignment of their satellite planes with respect to their obscuring disc, which makes sure the part that is most-distant from the best-fitting plane is effectively hidden in the obscuration volume. The other obvious difference between the samples is that we use the Millennium II simulation scaled to cosmological parameters from WMAP7 (instead of WMAP1), while the semi-analytic modelling is almost identical \citep{Guo2013}.
Thus, using only one of the single-criterion comparisons ($c/a$ or $r_{\mathrm{per}}$) it can already be concluded now that it is considerably less likely for a simulated satellite system to reproduce the observed flattening VPOS than claimed by W13, if the MW discs are oriented randomly.

As discussed in \citet{Pawlowski2013a}, a more concentrated satellite distribution will naturally have a smaller rms height $r_{\mathrm{per}}$\ on average. This can be seen in Fig. \ref{fig:wang:rparrper}, which plots $r_{\mathrm{per}}$\ against $r_{\mathrm{par}}$, the rms distance of satellites projected on to the best-fitting plane: small $r_{\mathrm{per}}$\ are preferentially present in satellite systems with small $r_{\mathrm{par}}$. Therefore, following the criteria of BB14, in order to resemble the spatial distribution of the MW satellites the simulated systems not only have to be narrow ($r_{\mathrm{per}}^{\mathrm{sim}} \leq r_{\mathrm{per}}^{\mathrm{obs}}$) but should also be at least as radially extended ($r_{\mathrm{par}}^{\mathrm{sim}} \geq r_{\mathrm{par}}^{\mathrm{obs}}$). This is the case for only 1.2 per cent of the systems. 

Requiring the relative shape of the satellite distribution to be at least as flattened as observed ($(c/a)^{\mathrm{sim}} \leq (c/a)^{\mathrm{obs}}$\ and $(b/a)^{\mathrm{sim}} \geq (b/a)^{\mathrm{obs}}$) results in a similar fraction of 1.6 per cent of the systems (see Fig. \ref{fig:wang:axratios}). 
If the satellite positions are made isotropic by randomizing their angular positions, the probability to find a system similar to the observed one is lower (0.3 and 0.6 per cent for the combined criteria, respectively).

\subsubsection{A co-orbiting VPOS}

Aside from the spatial distribution, a major feature of the VPOS is the co-orbiting sense of the satellite galaxies. This can be expressed by two parameters: the spherical standard distance of the eight best-aligned orbital poles, $\Delta_{\mathrm{std}}$, and the alignment of their average direction with the normal vector to the best-fitting satellite plane, $\theta_{\mathrm{VPOS}}$. Applying this test to the simulated satellite systems results in criteria 5 and 6 in Table \ref{tab:VPOSresults}. In 2.5 per cent of the simulated satellite systems eight of the 11 satellites have orbital poles that align to better than $\Delta_{\mathrm{std}}^{\mathrm{obs}} = 29.3^{\circ}$\ determined from observational data. This is the case for only 0.16 per cent in the randomized satellite sample, indicating that there is indeed some degree of coherence in the orbital direction of dark matter sub-haloes. These fractions are comparable to those reported in \citet{Pawlowski2013b}. For a random distribution of orbital poles they find a probability of 0.10 per cent that $\Delta_{\mathrm{std}}^{\mathrm{sim}} \leq \Delta_{\mathrm{std}}^{\mathrm{obs}}$, while for sub-haloes drawn at random from the Aquarius or Via Lactea simulations they find an average of 0.64 per cent and a maximum value of 1.67 per cent for the Aquarius halo C2. The analysis of \citet{Pawlowski2013b} does not exclude objects close to the MW plane, which explains why the fractions we find now are somewhat larger.
The average direction of the best-aligned orbital poles aligns with the normal direction of the best-fitting plane in 13.4 per cent of the simulated samples and in 10.1 per cent of the randomized ones. The agreement of these values indicates that there is only a weak tendency for the simulated galaxies to align with the plane defined by their positions.

To reproduce the situation of satellite galaxies co-orbiting in the VPOS, the simulated satellite orbital poles have to be both concentrated and close to the normal defined by the plane fit to their positions, i.e. they have to fulfil both criteria simultaneously. Only 1.0 per cent of the simulated satellite systems do so while such a situation is essentially non-existent for the randomized satellite distributions (upper limit 0.006 per cent).

\subsubsection{A thin and co-orbiting VPOS}

\begin{figure}
   \centering
   \includegraphics[width=85mm]{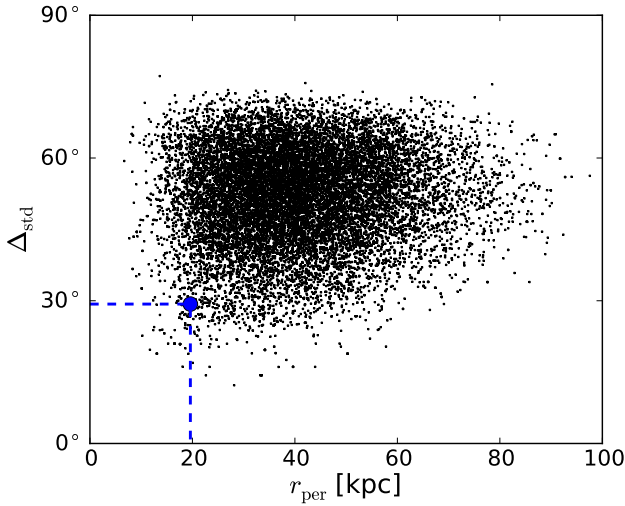}
   \caption{Orbital pole concentration $\Delta_{\mathrm{std}}$\ plotted against $r_{\mathrm{per}}$, the rms height of the satellite galaxies from the plane fitted to the positions of the 11 brightest satellites around each host. Shown are the results for the samples of 11 satellites. The black points represent the results for the Millennium II simulations, the blue dot the values for the observed satellites around the MW. To be as narrow and as coherently rotating as the MW satellite system, the simulations must lie in the region to the left and below the dashed lines. This is the case for only 0.35 per cent of all simulated satellite samples (criterion \#9 in Table \ref{tab:VPOSresults}). If the average orbital pole is simultaneously required to be as close to the plane normal as is the case for the observed VPOS, only 0.15 per cent of the simulated satellite systems can reproduce the observed situation (criterion \#14 in Table \ref{tab:VPOSresults}).
   }
              \label{fig:wang:rperpole}
\end{figure}

When considering the spatial distribution or the orbital pole distribution separately only about 1 per cent of the simulated satellite systems can reproduce the observed properties of the MW satellite system. However, to be in agreement with the VPOS both the position as well as the orbital poles have to be considered simultaneously. For both combinations of four criteria (either $r_{\mathrm{per}}$, $r_{\mathrm{par}}$, $\Delta_{\mathrm{std}}$\ and $\theta_{\mathrm{VPOS}}$\ or $c/a$, $b/a$, $\Delta_{\mathrm{std}}$\ and $\theta_{\mathrm{VPOS}}$), only 9 respectively 8 out of 16930 simulated satellite systems reproduce the observed situation, only about 0.05 per cent. This is equivalent to less than one of the 1693 host haloes. 
Even requiring that only three of the criteria are met simultaneously (ignoring either the requirement for the radial distribution of the satellites, $r_{\mathrm{par}}$, or the alignment of the average orbital pole with the plane normal, $\theta_{\mathrm{VPOS}}$) does not increase these numbers substantially (criteria 12 to 15 in Table \ref{tab:VPOSresults}). It is in fact already very unlikely that a simulated satellite systems happens to have a sufficiently concentrated distribution of orbital poles and a small rms height (0.35 per cent, criterion 9, see Fig. \ref{fig:wang:rperpole}) or axis ratio (0.28 per cent, criterion 10) at the same time.

\subsubsection{Including Canes Venatici I in the VPOS}

W13 arbitrarily excluded the MW satellite CVnI from their sample of observed galaxies. However, they also showed that as the number of simulated satellites in a system increases, the axis ratio $c/a$\ increases as well. Our results indicate that this trend has already an effect if only one additional satellite galaxy is included. The fraction of simulated systems fulfilling criterion 3 in Table \ref{tab:VPOSresults} changes from $P_{\mathrm{sim11}} = 3.4$\ per cent to $P_{\mathrm{sim12}} = 2.4$\ per cent. 
As discussed above, it is more likely to draw 8 closely aligned orbital poles out of a sample of size 12 instead of size 11. This is reflected in the slightly higher probabilities $P_{\mathrm{sim12}}$\ compared to $P_{\mathrm{sim11}}$\ to fulfil criterion 5 in Table \ref{tab:VPOSresults}. This probability and those based on it are upper limits and will drop if the orbital pole of CVnI turns out to be well-aligned with the others. In any case, the probabilities to reproduce the properties of the observed MW system are similar and close to zero for both sample sizes. Including CVnI in the analysis therefore does not change our main result and slightly reduces the likelihood for a simulated satellite system to resemble the observed VPOS.

\subsection{Summary}

Whether considering the 11 or 12 MW satellites with the largest baryonic mass, in both cases the observed thinness or flattening, orbital pole coherence and alignment of orbits with the best-fitting plane is reproduced by $\lesssim 0.2$\ per cent of the simulated satellite systems, and less than 0.06 per cent of all satellite systems have the four observed properties simultaneously. The observed MW satellite galaxies constituting the VPOS are simply too well-aligned in phase-space to be reproduced by sub-halo based satellite galaxies found in the Millennium II simulation. 

If additional satellite objects were to be considered, such as the faint satellite galaxies \citep{Kroupa2010}, globular clusters and stellar and gaseous streams in the MW halo \citep{Pawlowski2012a}, the probability to find the observed situation in these $\Lambda$CDM simulations would be even smaller. It is worth noticing that this analysis relies only on the assumption that all classical MW satellites, brighter than $M_{\mathrm{V}} \approx -8$\ and not obscured by our position within the stellar disk, have been discovered within a radius of 260~kpc. It does not rely on the distribution of rogue ultra-faint dwarf spheroidal galaxies that might be found outside the VPOS in the future.

The results of this section have been obtained by following the method of W13, except for the changes listed in Sect. \ref{sect:wang:method}. 
It would of course be interesting to know if these results can be reproduced if instead of the haloes with largest mass in baryons $M_{\mathrm{baryon}}$\ those with the highest V-band luminosity $L_V$\ or the highest dark matter mass $M_{\mathrm{peak}}$\ or the largest circular velocity $V_{\mathrm{peak}}$\ before infall would be selected, but his is beyond the scope of this paper and left for future studies. One would expect that the haloes with the largest $M_{\mathrm{baryon}}$\ have a high chance also to be among the ones with the highest $L_V$, $M_{\mathrm{peak}}$, $V_{\mathrm{peak}}$\ or other parameters of a similar type. This expectation is supported by the comparison for sub-haloes from the Aquarius simulation presented in W13. They find that a fraction of about 80 to 85 per cent of the $\approx 11$\ highest-ranking satellites are identical to those selected using semi-analytic modelling if the selection parameter is changed to $V_{\mathrm{peak}}$\ or $M_{\mathrm{peak}}$. Thus it appears likely that similar results will be obtained when using these different parameters.

Again it would be helpful to have higher-resolution simulations available such that no orphan galaxies have to be included in the analysis. Such studies are currently carried out.

%__________________________________________________________________

\section{The GPoA via cold stream accretion?}
\label{sect:goerdt}

\subsection{Previous method and its limitations}
\label{sect:goerdt:problems}

GB13\footnote{Even though that work in its present form is only a non-refereed arXiv preprint, it is being cited (including by published refereed papers) as indicative that cold stream accretion might explain the GPoA (e.g. \citealt{Shaya2013}, BB14, \citealt{Hoffmann2014}).} claimed that the GPoA can be understood if all satellite galaxies are accreted along cold streams that have been proposed as the dominant accretion mechanism for baryons at high redshift ($z \geq 2$) based on theoretical work and simulations \citep[e.g.][]{Dekel2009}.
GB13 assume that the satellite galaxies are accreted in narrow streams of cold gas, which they model as straight lines crossing the centre of the host galaxy. To determine how likely it is that a certain fraction of satellites lie within a narrow plane, they then determine how often a fraction of lines close to the fraction of M31 satellites in the GPoA (15 out of 27) of such randomly oriented lines are found in a common plane with a sufficiently small opening angle. Their main result is that it is very likely that two to four out of three to eight randomly oriented lines align within a narrow plane.

\begin{figure}
   \centering
   \includegraphics[width=85mm]{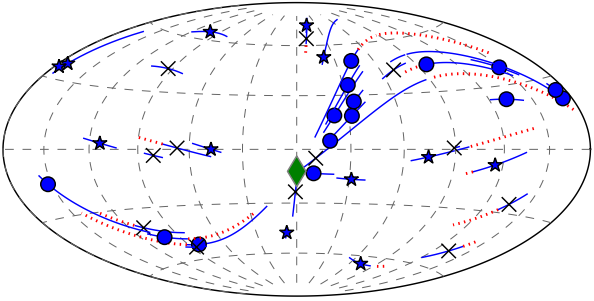}
   \caption{All-sky plot as seen from the centre of M31. The symbols indicate the infall directions of M31 satellites determined from their positions relative to M31 and their line-of-sight velocities, assuming them to be on perfectly radial orbits. Circles indicate satellites which belong to the GPoA as identified by \citet{Ibata2013}, stars belong to the other satellites in the \citet{Ibata2013} sample and crosses indicate the remaining known galaxies within 500\,kpc from M31.
   The lines indicate the position uncertainties of the infall directions corresponding to the $1\sigma$\ uncertainties in the heliocentric distances. The red dotted sections of lines indicate that the satellite would not be infalling from but receding from M31 in that direction, because its distance uncertainty makes it change from the side closer to the MW than M31 to the far side or vice versa. 
While several of the satellites belonging to the GPoA have similar infall directions, close to the position of the MW indicated by the green diamond, most of the satellite galaxy infall directions are widely separated. This is in contrast with the scenario assumed in GB13, in which they all share a few radial orbits. The assumption of all satellite being accreted radially along a few cold streams is not valid if one intends to explain the GPoA with cold stream accretion.
   }
              \label{fig:goerdt:ASP}
    \end{figure}

\begin{figure}
   \centering
   \includegraphics[width=85mm]{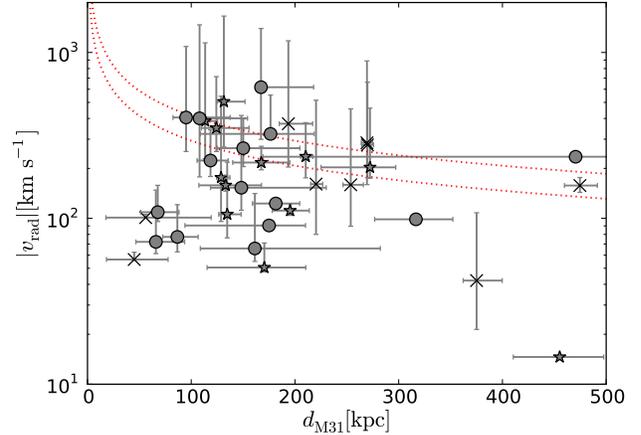}
   \caption{Median absolute velocities of the M31 satellite galaxies assuming that they are on radial orbits with respect to M31, plotted against their median distances from M31. The error bars indicate the $1\sigma$\ quantiles of the respective quantities derived from varying the satellite and M31 positions within their observational uncertainties. Symbols are the same as in Fig. \ref{fig:goerdt:ASP}. Several galaxies have velocities that are likely exceeding the escape velocities of point masses of 1.0 and $2.0 \times 10^{12} \mathrm{M}_{\sun}$, marked as the dotted red lines. This makes it very unlikely that all satellites have been accreted along a few common cold streams at $z \geq 2$, as the unbound ones would have left the vicinity of M31 in the meantime. For comparison it shall be mentioned that all but one of the MW satellites for which proper motions have been estimated have dominating tangential velocity components and that they all have radial velocities well below $200\,\mathrm{km\,s}^{-1}$\ \citep[e.g.][]{Fouquet2012,Pawlowski2013b,Barber2014}.
   }
              \label{fig:goerdt:vrad}
    \end{figure}

There are a number of limitations to the analysis of GB13 and its accord with the observed situation. The most important issues are:

\begin{enumerate}
  \item The validity and importance of the cold stream accretion mode itself as well as its connection to satellite accretion has not yet been observationally confirmed. In particular, it is known that dwarf spheroidal galaxies have higher than Newtonian internal velocity dispersions, which in the standard context means that they have to be embedded in dark matter haloes. How these sub-haloes share exactly the dynamics of cold gas along these thin streams in order to then accrete and form satellite galaxies within dark matter haloes still has to be demonstrated.
  In addition, recent numerical simulations suggest that the cold flow accretion mode might have been over-estimated and could therefore to a certain extend be numerical artefacts of the simulation codes used \citep{Nelson2013}.
  \item Cold stream accretion is believed to be most important at high redshifts ($z \gtrsim 2$) \citep[e.g.][]{Dekel2006,Ocvirk2008,Keres2009}. GB13 therefore implicitly assume that all of the satellite galaxies have been acquired at about that time. This is in conflict with the findings by \citet{Gao2004}, according to which about 90 per cent of the present-day sub-haloes in cosmological simulations have been accreted after $z = 1$. The survival mass fraction for haloes accreted at z = 2 is just 2 per cent. Tidal effects would have strongly affected these satellites, in particular if they were moving on radial orbits. This would make the satellites pass through the centre of M31, where they would be tidally disrupted and where they would suffer orbital decay due to dynamical friction \citep{Angus2011}. The scenario presented by GB13 therefore requires the ad-hoc assumption that satellites accreted via cold streams are much more stable than usual dark matter sub-haloes.
  \item The analysis of GB13 implies that the M31 satellites share a few radial orbits, because they were accreted radially from only three to eight different directions. However, they do not test whether this is consistent with the observational data. We have done so using two different methods: 
\begin{itemize}
     \item The scenario of GB13 implies that the M31 satellites all originate from no more than eight different directions, as this is the maximum number of streams in their analysis. Because the satellites are supposed to move on radial orbits, their positions as seen from M31 today should unambiguously identify their orbits (which in this scenario are the lines connecting the centre of M31 with the satellites). We can also deduce a satellite's orbital phase from its observed line-of-sight velocity relative to M31's line-of-sight velocity. If a satellite on a radial orbit is closer to the MW than M31 and is approaching us faster than M31, it has to be receding from M31. Thus, its infall direction has to be on the opposite side of M31, so in an all-sky-plot of M31 the mirrored position of the satellite identifies the infall direction. Deducing the original infall direction in this way for all galaxies within 500\,kpc of M31, we arrive at the all-sky-plot of putative infall directions shown in Fig. \ref{fig:goerdt:ASP}. If the assumptions of GB13 were correct, the infall directions of the satellite galaxies should cluster very close to a small number of common directions. This is not the case, demonstrating that the satellite galaxies do not share a few radial orbits, and therefore can not have been accreted along a few radial streams.
     \item As a second test we can make use of the simplicity of the orbital geometry implied by GB13. 
  Assuming a radial orbit, a satellite's measured line-of-sight velocity (in the M31 rest-frame) is just the projection of its radial velocity towards (or away from) M31 on to the line-of-sight. Therefore, we can determine the radial velocity of the satellite from its position relative to M31 and its line-of-sight velocity. As the uncertain satellite position relative to M31 has significant influence on the resulting radial velocity, we use a Monte-Carlo method to vary the positions of the satellite and of M31 within their uncertainties. The resulting absolute radial velocities $|v_{\mathrm{rad}}|$\ of the M31 satellites are plotted against their distances $d_{\mathrm{M31}}$\ from M31 in Fig. \ref{fig:goerdt:vrad}. Some of the satellite galaxies, both within the GPoA and outside, would have very high absolute velocities of more than $300\,\mathrm{km\,s}^{-1}$ under that assumption. They are likely exceeding the escape velocities from point masses of 1.0 and $2.0 \times 10^{12} \mathrm{M}_{\sun}$, marked by the two red dotted lines\footnote{Treating M31 as a point mass is a conservative approximation as well, as for a given total mass, the mass distribution with the highest escape velocity is a point mass.}. Such satellites would therefore not have remained bound to M31 if they were accreted on radial orbits at $z \geq 2$ as considered by GB13, in particular as the galaxy's mass was probably considerably smaller at that time. The fact that the satellites are still close to M31 today therefore requires them to move on non-radial orbits, such that they are in conflict with the basic assumption of GB13. 
\end{itemize}
  \item Assuming that the MW and M31 satellites originate from cosmological or cold stream accretion, there is no reason to consider that the MW has acquired its satellite galaxies in a completely different way than M31. Of the 11 satellite galaxies for which proper motion measurements are available, all but one have a dominating tangential velocity component \citep[see e.g. fig. 3 in ][]{Pawlowski2013b}, and they all have radial velocities well below $200\,\mathrm{km\,s}^{-1}$\ \citep[e.g.][]{Fouquet2012,Barber2014}. The hypothesis of radial orbits is therefore not supported by the observed velocities of the MW satellites.
  \item Cold streams are not expected to be perfectly radial. For example \citet{Wang2014} assume an impact parameter of 30\,kpc. The angular momentum of streams brought into a satellite system is non-negligible. Therefore, the stream material and possible satellites accreted along it will share a common orbital plane, but not be on a radial orbit. Furthermore, we have to consider the observed co-orbiting sense of the GPoA, i.e. that 13 out of the 15 satellites have the same orbital sense \citep{Ibata2013}. This has been ignored in the analysis of GB13. The preference of the GPoA satellites to co-orbit is an integral part of the peculiarity of that structure and further emphasises the importance of incorporating the orbital direction of accreted satellite galaxies in the analysis.
  \item At least one cold stream will feed the formation of the central galactic disc. Gas clumps brought in with the stream merge with the central galaxy and the stream will not be available to bring in satellite galaxies \citep{Dekel2009}.
  \item
  Cosmological zoom-in simulations of the cosmic web \citep[e.g.][]{Ocvirk2008,Ceverino2010} from which the cold accretion scenario of galaxy mass assembly has been implied \citet{Dekel2009} show that clumpy substructures develop within the gas filaments. Depending on the spatial and mass resolutions of the simulations, the clump masses can be understood as representing high-mass satellites. GB13 follow this interpretation. The separation of these density structures in simulations varies between 25 and 100 kpc. If one such cold filamentary stream should have caused the accretion of the satellite system of M31, the accumulation of 50 substructures (less than twice the number of the observed satellites to allow for the accretion of clumps by the parent galaxy) separated by 50\,kpc with an infall velocity of 200\,km\,s$^{-1}$\ would take about 12.5\,Gyr. Since the cold stream infall would be preferably directed towards the LG's centre of mass, the MW and M31 would move along their orbits and by this change the impact parameter and direction of the cold streams, invalidating the assumptions of GB13. The accretion of all satellites along one or a small number of streams is therefore unrealistic.
\end{enumerate}

Many of these limitations and inconsistencies with the observed situation are due to the over-simplified assumption that the satellite galaxies are confined to a few perfectly radial orbits, which GB13 have modelled as straight lines. A number of these limitations have already been pointed out by GB13, but they did not account for them in their analysis. In the following, 
we modify their analysis by assuming that each stream defines an orbital plane instead of a radial orbital line. This also allows us to make use of the observational information regarding the co-orbiting satellites in the GPoA.

\subsection{New method}

\begin{figure*}
   \centering
   \includegraphics[width=58mm]{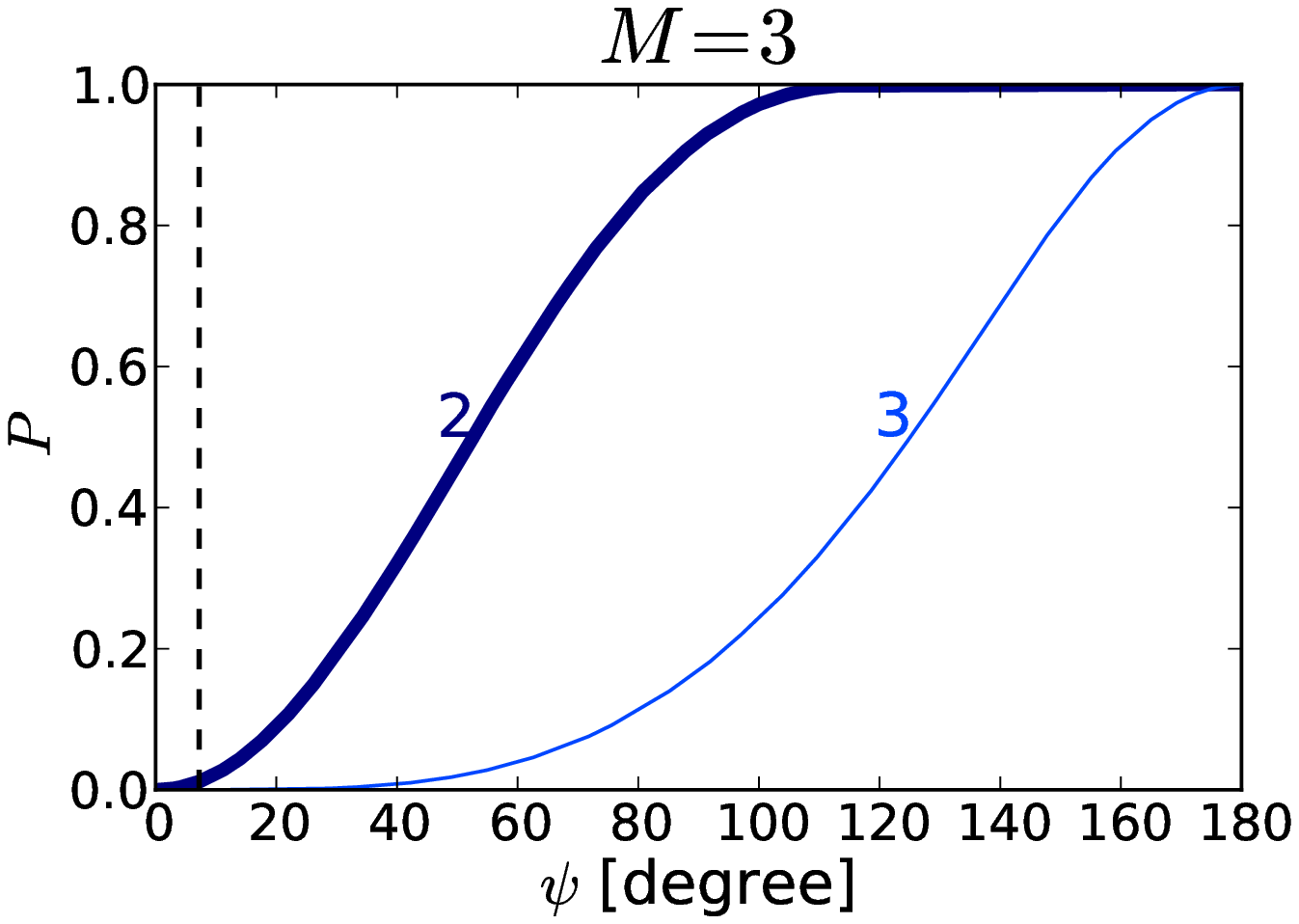}
   \includegraphics[width=58mm]{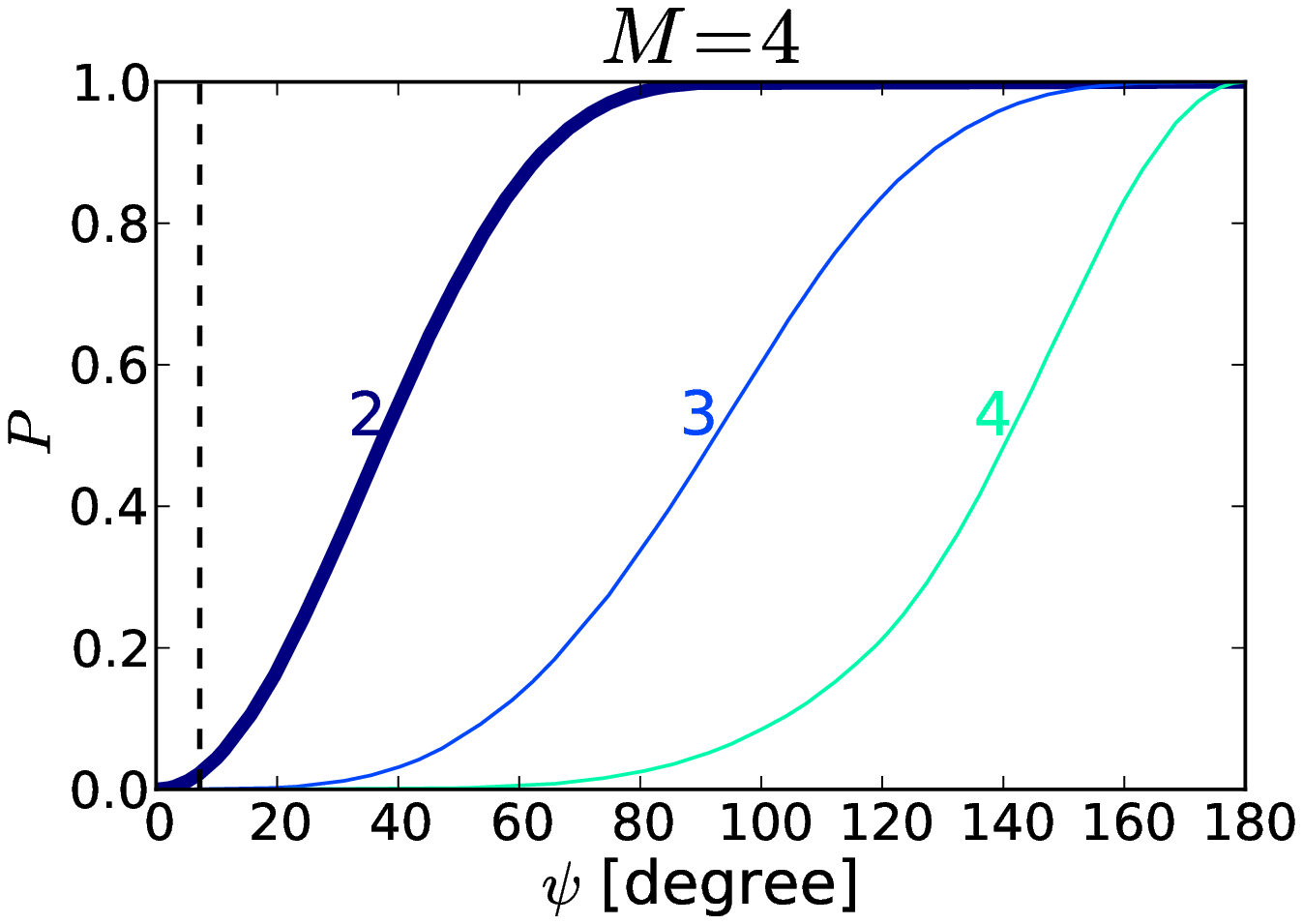}
   \includegraphics[width=58mm]{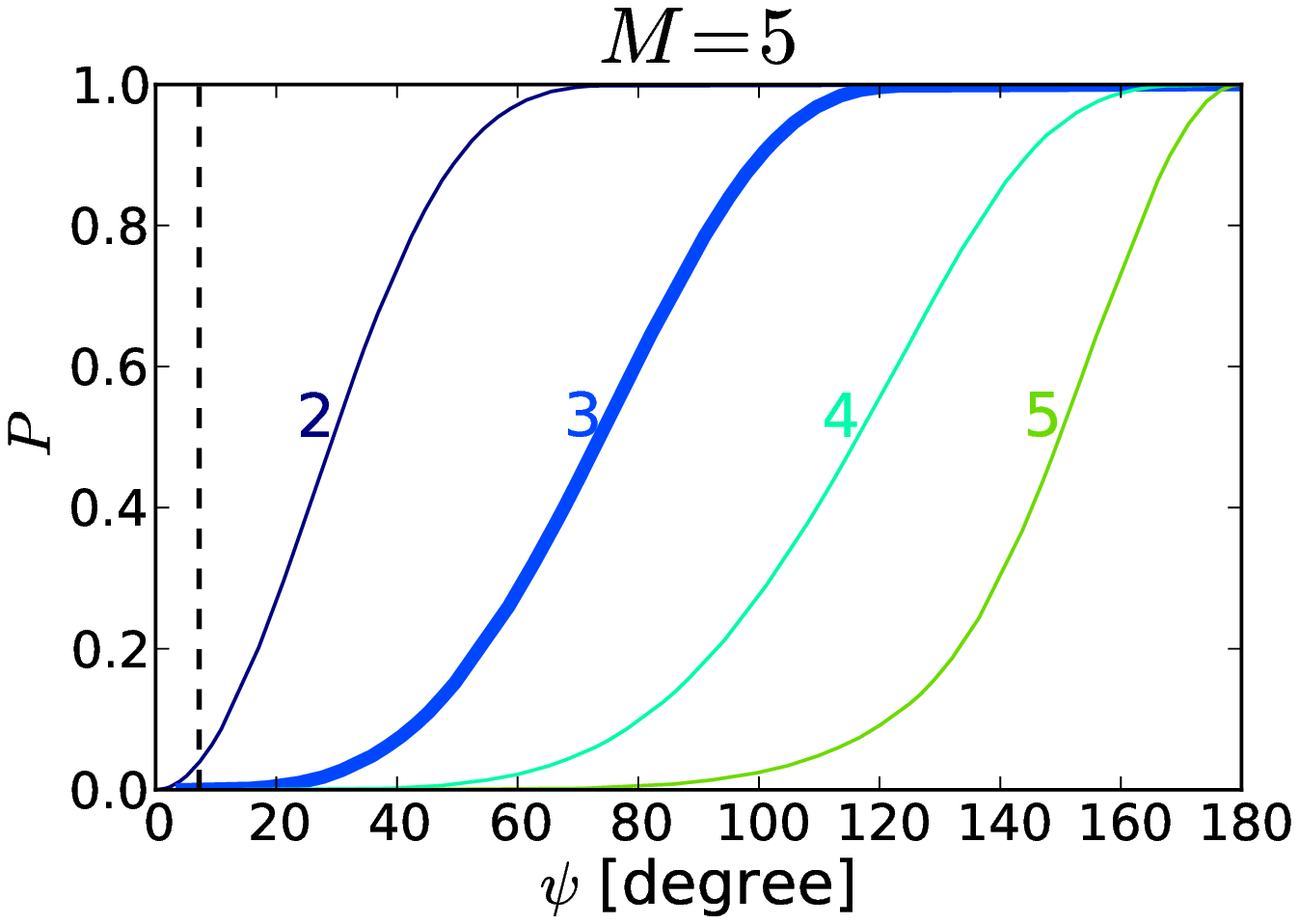}
   \includegraphics[width=58mm]{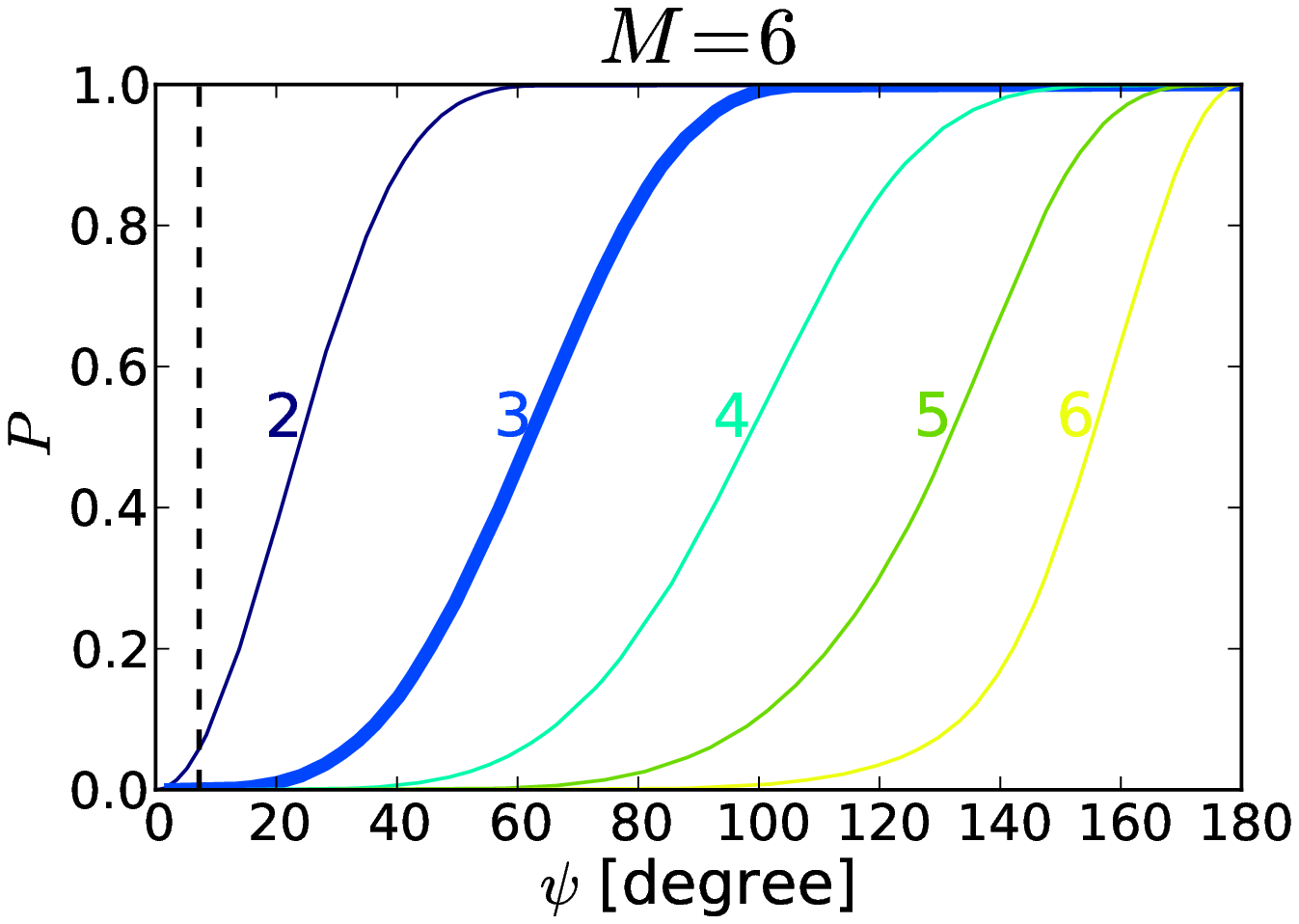}
   \includegraphics[width=58mm]{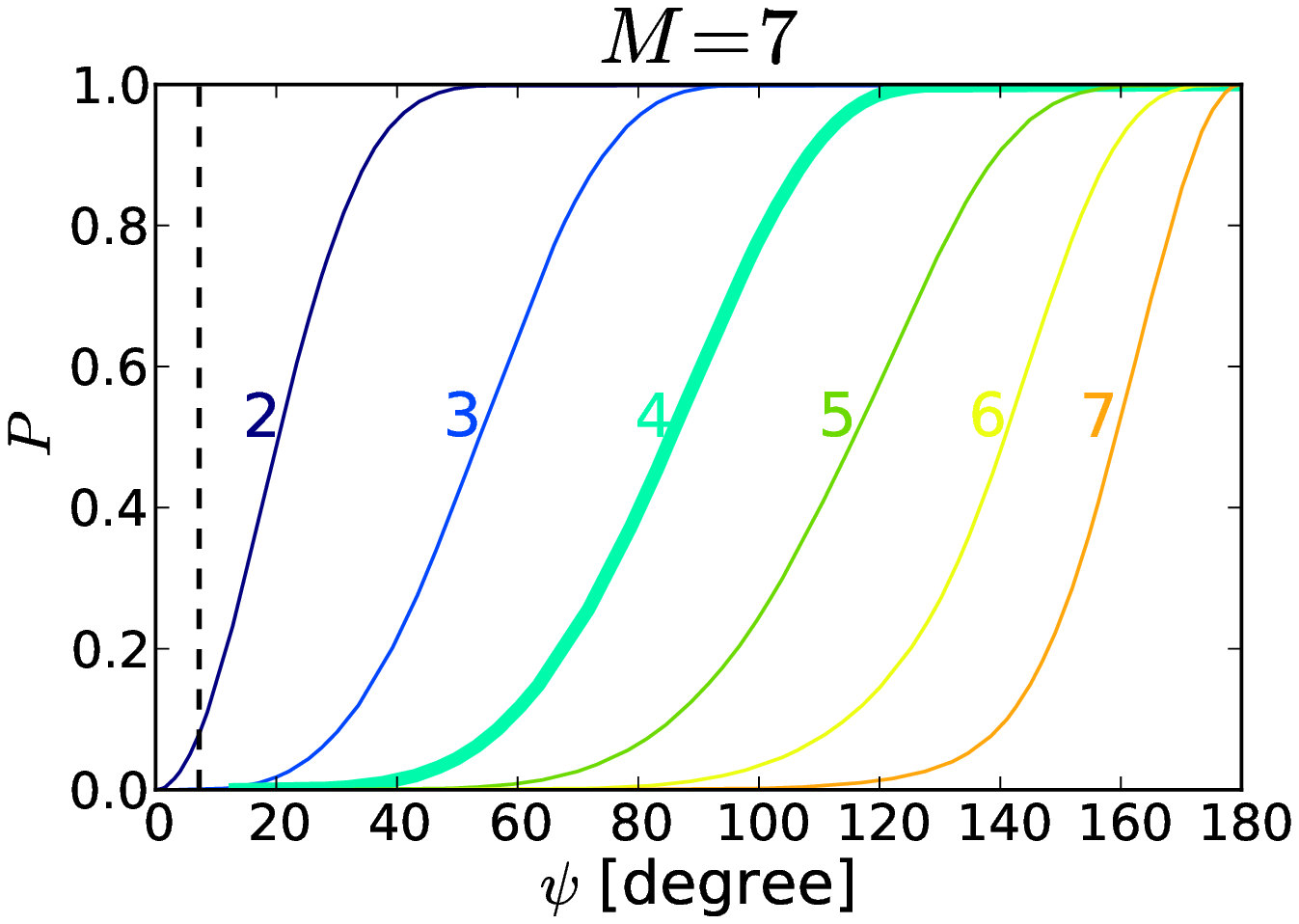}
   \includegraphics[width=58mm]{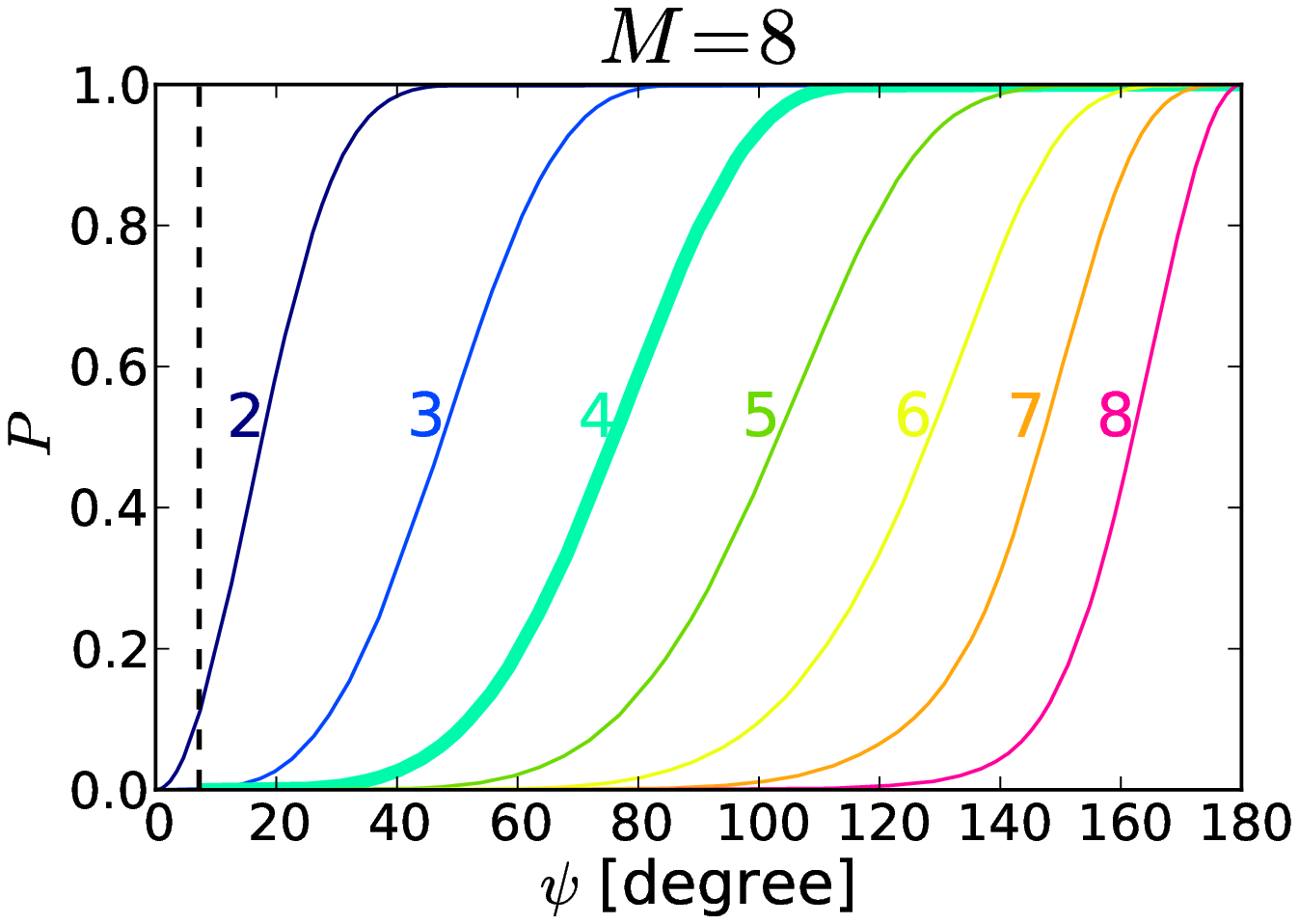}
   \caption{Cumulative probabilities to find at least $n$\ out of $M$\ randomly oriented stream poles within an angle $\psi$\ or less of each other. Each panel presents the results of a different value for $M$, the total number of streams feeding the host. In each panel, the cumulative probabilities of the different $n$\ are color coded, the number $n_{\mathrm{GPoA}}(M) = \lceil 13/27 \times M \rceil$\ corresponding to the observed one is marked as a thicker line. The dashed line indicates the opening angle of the observed GPoA around M31, $\psi_{\mathrm{GPoA}} = 7^{\circ}.2$.
   In contrast to the findings of GB13, who assumed that all M31 satellites move along lines passing through M31's centre, our results demonstrate that it is very unlikely to find $\geq 13/27$\ of the satellite galaxies within a thin, co-orbiting plane consistent with the observed one if the streams are described as orbital planes.
   }
              \label{fig:goerdtlikelihoods}
    \end{figure*}

We adopt the hypothesis of GB13 that (i) all satellites are accreted via cold streams, (ii) 3-8 streams are feeding the host halo, (iii) the streams are randomly oriented and (iv) each stream is loaded with an equal amount of satellites. Our analysis still makes the (unreasonable) assumption that the origin of the streams does not change, which would spread out the planes. This assumption therefore over-estimates the likelihood to find a sufficiently thin satellite plane. As discussed before it is in conflict with cosmological simulations and dynamical friction to assume that all present day satellites have been accreted before $z = 2$, when cold streams might have been an important accretion mechanism. We will therefore discuss in Sect. \ref{sect:goerdt:results} how the situation changes when assuming that half of the satellites have been accreted from directions unrelated to the streams.

Instead of modelling the cold streams as lines crossing the centre of the host galaxy, which assumes that all satellite galaxies are accreted along and remain on perfectly straight radial orbits, it is better to describe the streams as orbital planes. A stream flowing not exactly towards a host galaxy but having a (small) impact parameter will result in a common orbital plane for the stream material. The exact impact parameter is not important in this description. It only has to be assumed that the streams will stay within their original orbital plane. Precession effects in non-spherically-symmetric potentials would invalidate this assumption and result in a thickening of the satellite planes (Klarmann et al. in prep.), such that this assumption again results in an over-estimation of the probability to find sufficiently thin satellite planes.
In our analysis, each stream is described by its pole, which is the direction of the angular momentum of the stream material. This stream pole is a normal vector to the stream's orbital plane, but also contains information about the orbital sense of accreted streams\footnote{In fact we would have to require at least one stream to feed the growth of M31's galactic disc, but refrain from incorporating this possibility for the sake of simplicity. Later merger events might also change the host galaxy's spin direction.}.

For each realization, we generate $M$\ randomly oriented stream poles by drawing them from an isotropic distribution, following the recipe of GB13. We then run through all possible combinations of $n = [2, 3..., M]$\ streams and measure the largest angle between all the stream poles in each of the $\frac{M!}{(M-n)!\,n!}$\ possible combinations for each of the $M-1$\ different sub-sample sizes $n$. For each combination of streams we determine the maximum angle between the stream poles. This is the opening angle of that stream combination. We then record the smallest opening angle, $\psi$, among all stream combinations of a given sub-sample size $n$.
This is equivalent to the opening-angle of the best-aligned group of streams. As we use the stream poles and not axial normal vectors, this assures that the streams are co-orbiting if $\psi < 90^{\circ}$. For each number of streams $M = [3,4,...,7,8]$ we generate $10^5$\ random realizations.

For the GPoA, GB13 use an opening angle of $3.6^{\circ}$. We double this value in our analysis. Thus, all sub-samples with $\psi_{\mathrm{GPoA}} \leq 7.2^{\circ}$\ are considered to result in planes as narrow as the observed GPoA. From the 27 satellite galaxies in the sample of \citet{Ibata2013}, 15 are aligned in the GPoA but only 13 of these co-orbit. As we, in contrast to the analysis of GB13, consider the orbital sense of the accreted satellites, we have to find $n_{\mathrm{GPoA}} = 13/27$\ of the $M$\ planes to be aligned to $\psi \leq \psi_{\mathrm{GPoA}}$\ to reproduce this observed situation (which assumes that all streams bring in the same number of satellites and ignores the two aligned but counter-orbiting galaxies in the GPoA). In determining the appropriate $n_{\mathrm{GPoA}}$\ for a given number of $M$\ streams we furthermore have to round up to the next integer, i.e. $n_{\mathrm{GPoA}}(M) = \lceil 13/27 \times M \rceil$. Rounding down would not result in a sufficient number of satellites within the GPoA, while rounding up allows some of the satellites to have been scattered out of the plane.

\subsection{Results}
\label{sect:goerdt:results}

\begin{table}
  \caption{Likelihoods to find a sufficient number of aligned stream orbital planes.}
  \label{tab:goerdt}
  \begin{center}
  \begin{tabular}{lccc}
  \hline
  $M$ & $n_{\mathrm{GPoA}}$ & $P_{\mathrm{GPoA}}$ [\%] & $P(n = n_{\mathrm{GPoA}} -1)$ [\%]\\
 \hline
3 & 2 & 1.157 & 100.0 \\
4 & 2 & 2.227 & 100.0 \\
5 & 3 & 0.006 & 3.874\\
6 & 3 & 0.015 & 5.830\\
7 & 4 & 0.000 & 0.024\\
8 & 5 & 0.000 & 0.035\\
\hline
\end{tabular}
 \end{center}
 \small \medskip
  Probabilities $P_{\mathrm{GPoA}}$\ to find $n_{\mathrm{GPoA}}$\ out of $M$ randomly oriented stream planes to be aligned in a co-orbiting orientation. $P(n = n_{\mathrm{GPoA}} -1)$\ gives the probability to find $n_{\mathrm{GPoA}} - 1$\ aligned planes, which however is not sufficient to explain the observed number of co-orbiting satellites in the GPoA.
\end{table}    
    
Fig. \ref{fig:goerdtlikelihoods} compiles the results of our analysis. Each panel shows the cumulative distribution function of the best-aligned $n$\ stream orbit planes for a different number of streams $M$ plotted against the opening angle $\psi$\ of the best-aligned combination of planes for a total of $10^5$\ realizations.

A situation in which a sufficient number $n_{\mathrm{GPoA}}$\ of streams aligns in a co-orbiting plane with an opening angle smaller than or equal to that of the GPoA is very unlikely (see the probabilities listed in Table \ref{tab:goerdt}). The largest probability $P_{\mathrm{GPoA}} = 2.28$\ per cent that $n_{\mathrm{GPoA}}(M=4) = 2$\ streams align to at least $7.2^{\circ}$ is found for $M = 4$\ streams.

Overall, an alignment of a sufficient number of streams is essentially excluded for $M \geq 5$\ as the probabilities $P_{\mathrm{GPoA}}$\ remain below 0.015 per cent in these cases. However, according to simulations it is most likely that galaxies are fed by a lower number of streams. Nevertheless, a situation comparable to the observed GPoA occurs in only one out of 50 galaxies, if assuming that all satellites are accreted exclusively in streams (see Sect. \ref{sect:goerdt:problems} for why this is unrealistic). 

While the likelihoods to have fewer streams than necessary to align are higher (right column in Table \ref{tab:goerdt}), this is insufficient to reproduce the number of co-orbiting satellites in the GPoA, as we have to require that \textit{at least} 13 out of 27 satellites are in a common plane. Satellites can be scattered out of the plane later (as was also argued by GB13), but it is less likely to add objects to the co-orbiting plane. Thus it is incorrect to include cases where the fraction of aligned planes is smaller than $n_{\mathrm{GPoA}}(M) = \lceil 13/27 \times M \rceil$.

As discussed in Sect. \ref{sect:goerdt:problems} it is expected that additional satellite galaxies have been accreted after the cold stream era. The addition of such satellites that fall in unrelated to the cold streams requires more objects to be in the common plane initially if the GPoA is to be explained by the cold stream scenario. If we assume for simplicity that 50 per cent of all satellites ($0.5 \times 27 = 13.5$) have been accreted individually\footnote{Note that this is a very low fraction given that, according to \citet{Gao2004}, 90 per cent of the sub-haloes that survived until today have been accreted after $z = 1$, whereas simulations indicate that cold stream are dominant only before a redshift of $z = 2$.} and that these follow (to first order) an isotropic distribution. Then only half of the M31 satellites would have been accreted in streams. This in turn requires that essentially all stream planes had to align initially. The reason is that, if drawn from isotropy, only $\sin(3.6^{\circ}) = 6.3$\,per cent of isotropically distributed satellites will end up within the GPoA., i.e. about one out of 13.5. As we also consider the co-rotating trend of the satellites, this number has to be divided by 2 once more, because there is only a 50 per cent chance that a satellite with a random direction of motion follows the common velocity trend. Thus, we can assume that on average 0.5 satellites out of the 13 co-orbiting in the GPoA are from the random distribution, while the remaining 12.5 have to originate from the alignment of cold stream orbital planes. As we assume that 50 per cent of the satellites (= 13.5) are accreted via cold streams, at least 12.5 out of these 13.5 had to align initially. Thus, essentially all cold stream planes had to be aligned. As can be seen in Fig. \ref{fig:goerdtlikelihoods}, the probability for this to happen is negligible for all assumed numbers of streams.

\subsection{Summary}
\label{sect:goerdt:discuss}

Given these results, we disagree with the conclusions of GB13. We have shown that with the simple but well motivated change from radial to planar satellite orbits we are already unable to reproduce their results. A sufficient alignment of streams occurs rarely (in at most $\approx 2$\ per cent of the systems with four streams, less in all other configurations). This finding stands in contrast to the results by GB13 who report probabilities as high as 50 to 100 per cent.

%__________________________________________________________________

\section{Conclusion}
\label{sect:conclusion}

We have searched for co-orbiting planes of satellite galaxies similar to the GPoA around M31 and the VPOS around the MW within satellite systems based on the Millennium II cosmological simulation. Our analysis was motivated by and is similar to the works by BB14 and W13. However, we have identified several limitations in their sample selection and methods, which we have addressed in our analysis.

Using the same simulation data, we find different results than BB14 concerning the frequency of the satellite plane around M31. A GPoA-like plane, i.e. a satellite structure which is sufficiently narrow, radially extended and contains at least 13 satellite galaxies with the same orbital sense, is found with a probability of only $P_{\mathrm{GPoA}} = 0.04$\ to 0.17 per cent, depending on the adopted criteria and assumed distance to Andromeda XXVII. This has to be contrasted to the 2 per cent found by BB14, who only required two of the three properties to be similar to the observed situation. Our analysis of their data therefore disputes their conclusion, according to which ``[...] Millennium II haloes with 13 or more corotating satellites orbiting in a thin plane are not rare [...]'' and ``[...] the existence of a [GPoA] is not in conflict with the standard cosmological framework''. The GPoA is extremely unlikely to be found in a dark-matter based $\Lambda$CDM universe within the limitations of current semi-analytic galaxy formation models. Its existence around M31 therefore poses a serious challenge to that and similar cosmological models of galaxy formation. 
These conclusions are supported and further strengthened by the analysis of the Millennium II simulation by  \citet{Ibata2014}, who apply the exact PAndAS survey area to the simulated satellite systems and require the satellite planes to be seen in the same edge-on orientation as the observed GPoA. Using this very different method, they find that satellite planes sharing the same parameters as the GPoA are extremely unlikely among simulated satellite systems (0.04 per cent).

We have also searched for satellite planes similar to the VPOS around the MW in the same dataset, incorporating the effect of randomly oriented obscuring MW discs. This analysis is comparable to the one conducted by W13. We are unable to reproduce their results on the frequency of satellite systems similarly flattened as observed, finding a 2 to 3 times lower frequency using only one criterion.To be comparable with the observed VPOS, a satellite system furthermore has to be not only as flattened as the 11 classical MW satellites, but also sufficiently radially extended, and coherently orbiting within the best-fitting plane. Only a fraction $P_{\mathrm{VPOS}} \leq 0.06$\ per cent of the simulated satellite systems fulfil these requirements simultaneously. In contrast to the conclusion by W13, our results evidence that the observed phase-space correlated satellite galaxy system of the MW alone already poses a serious challenge to the $\Lambda$CDM model, if it is well represented by the analysed Millennium II simulation in combination with semi-analytic modelling.

Another attempt to reconcile the observed GPoA with cosmological expectations has been made by GB13, who argue that the accretion of satellite galaxies along cold gas streams naturally results in planar satellite distributions. Their model of satellites moving on perfectly radial orbits along cold streams is an oversimplification, is in conflict with the observed M31 satellite galaxies, and is unable to produce the predominantly co-orbiting GPoA. 
We have adjusted their analysis by requiring all satellites accreted along a cold stream to remain in the same orbital plane, instead of on perfectly radial orbits, i.e. straight lines. With this modification we are unable to reproduce their results because it is much less likely that a sufficient number of randomly oriented planes align to the degree required to be consistent with the narrow, co-orbiting GPoA. Assuming further that only half of the satellite galaxies have been accreted along streams results in a negligible likelihood to obtain a structure as thin as the GPoA.

Both satellite galaxy systems for which we have sufficient observational data, that of the MW and of M31, are therefore in conflict with $\Lambda$CDM expectations derived from dark-matter-only simulations 
combined with a semi-analytic galaxy formation model. The reason for this can in principle either indicate a failure of the semi-analytic models on dwarf galaxy scale, missing physical processes in the simulation, or of the underlying cosmology \citep[see also][]{Kroupa2012a}.

The GPoA and the VPOS are both similarly unlikely to be found in the same satellite systems based on the Millennium II simulation, such that both structures can be considered to be of similar significance. As we have based the GPoA and the VPOS analyses on the same set of host galaxies, we are now also able to determine the $\Lambda$CDM-likelihood that both systems contain such structures.

Formally, taking the most favourable value for the GPoA of $P_{\mathrm{GPoA}} = 0.0017$ the probability of having a pair of two galaxies displaying one GPoA \textit{and} one VPOS, \textit{in any order}, is $P = 2 \times P_{\mathrm{GPoA}} \times P_{\mathrm{VPOS}} = 2.04 \times 10^{-6}$.

Nevertheless, one could argue that any individual configuration of satellite systems of two galaxies, if looked at with too much scrutiny, will have almost zero probability. To avoid this and be more conservative, we can consider that, since the two satellite planes have different properties (most importantly the GPoA consists of only a subset of the M31 satellites whereas almost all MW satellites are situated in the VPOS), the likelihood to have {\it either} a GPoA {\it or} a VPOS around any {\it one} random host galaxy is the sum of the individual probabilities\footnote{Some of the VPOS-like satellite systems might simultaneously be GPoA-like, thus the sum is only an upper limit.}:
\beq
P_{\mathrm{sum}} \leq P_{\mathrm{VPOS}} + P_{\mathrm{GPoA}} = 2.3 \times 10^{-3}. \nonumber
\eeq
This situation happens precisely twice in the two major galaxies of the Local Group, which has a probability
\beq
P_{\mathrm{LG}} = P_{\mathrm{sum}}^2 \leq 5.29 \times 10^{-6}. \nonumber
\eeq
The chance to find the VPOS- and/or GPoA-like satellite planes in the LG is therefore extremely small if all satellites are primordial dwarf galaxies in a $\Lambda$CDM universe.

Our work challenges recent claims that the observed satellite galaxy structures are consistent with the $\Lambda$CDM model. Co-orbiting planes of satellite galaxies such as the GPoA around M31 and the VPOS around the MW are still in conflict with satellite galaxy distributions derived from $\Lambda$CDM simulations. In fact, our results indicate that due to improved knowledge of the phase-space correlation of the systems they are even more in conflict with those than before.

%__________________________________________________________________

\section*{Acknowledgements}
We would like to thank Henning Bahl and Holger Baumgardt for providing the Millennium II data used in their study and for providing additional information on their analysis. We furthermore thank them for discussing their initial results on the MW and M31 satellite systems with MSP and PK in Summer 2013. We also thank Tobias Goerdt for discussing his work with GH.
MSP, PK, FL and HJ acknowledge financial support from DAAD/Go8 travel grant 56265912.
HJ also acknowledges the financial support from the Australian Research Council through the Discovery Project grant DP120100475.
DM was supported by the National Science Foundation under grant no. AST 1211602 and by the National Aeronautics and Space Administration under grant no. NNX13AG92G.
JD acknowledges funding through FONDECYT grant no.3140146.
DAF acknowledges the financial support from the Australian Research Council through the Discovery Project grant DP130100388.

\footnotesize{
\bibliographystyle{mn2e}
\bibliography{references}
}

\label{lastpage}

\end{document}